\pdfoutput=1
\newcommand*{\ATLASLATEXPATH}{}
\documentclass[PAPER, atlasdraft=true, texlive=2016, UKenglish, cernpreprint]{\ATLASLATEXPATH atlasdoc}
 
\usepackage{\ATLASLATEXPATH atlaspackage}
\usepackage{multirow}
\usepackage{rotating}
\usepackage{siunitx}
\usepackage{\ATLASLATEXPATH atlasbiblatex}
 
\usepackage{\ATLASLATEXPATH atlasphysics}
 
\graphicspath{{logos/}{figures/}}
 
\usepackage{ANA-DAPR-2018-01-PAPER-defs}

\AtlasTitle{ATLAS data quality operations and performance for 2015--2018 data-taking}
 
\AtlasAbstract{
The ATLAS detector at the Large Hadron Collider reads out particle collision data from over 100 million electronic channels at a rate of approximately $100$~kHz,
with a recording rate for physics events of approximately 1~kHz.
Before being certified for physics analysis at computer centres worldwide, the data must be scrutinised to ensure they are
clean from any hardware or software related issues that may compromise their integrity.
Prompt identification of these issues permits fast action to investigate, correct and potentially prevent future such problems that could render the data unusable.
This is achieved through the monitoring of detector-level quantities and reconstructed collision event characteristics at key stages of the data processing chain.
This paper presents the monitoring and assessment procedures in place at ATLAS during 2015--2018 data-taking.
Through the continuous improvement of operational procedures,
ATLAS achieved a high data quality efficiency, with 95.6\% of the recorded proton-proton collision data collected at $\sqrt{s}=13$~\ensuremath{\text{Te\kern -0.1em V}}\ certified for physics analysis.
}
 
\author{The ATLAS Collaboration}
 
\AtlasRefCode{DAPR-2018-01}
 
\AtlasJournal{JINST}
\AtlasJournalRef{JINST 15 (2020) P04003}
\AtlasDOI{10.1088/1748-0221/15/04/P04003}
 
\PreprintIdNumber{CERN-EP-2019-207}

\AtlasCoverSupportingNote{ATLAS Data Quality in Run 2}{https://cds.cern.ch/record/2663169}

\AtlasCoverAnalysisTeam{L.~Adam, M.~Aoki, E.~Baldin, E.~Bergeaas Kuutmann, C.~Bernius, R.~Bielski, M.~Biglietti, A.~Bingul, A.~Bruni, B.~Burghgrave
% ...
}

\AtlasCoverEdBoardMember{L.~Fiorini (chair), J.~Leveque, J.~H.~Stelzer}

\AtlasCoverEgroupEditors{atlas-DAPR-2018-01-editors@cern.ch}
 
\AtlasCoverEgroupEdBoard{atlas-DAPR-2018-01-editorial-board@cern.ch}

\hypersetup{pdftitle={ATLAS document},pdfauthor={The ATLAS Collaboration}}
 
\begin{document}
 
\maketitle
 
\tableofcontents

\section{Introduction}
\label{sec:intro}
 
The ATLAS detector~\cite{PERF-2007-01} at the CERN Large Hadron Collider (LHC)~\cite{Evans:2008zzb} is a multipurpose particle detector,
optimised to exploit the full physics potential of high-energy proton--proton ($pp$) and heavy-ion collisions.
This includes the study of the Higgs boson, precise measurements of Standard Model processes, searches for rare and new phenomena,
as well as the study of the properties and states of matter at high temperatures.
ATLAS has over 100 million electronic channels, which provide read-out for multiple, distinct detector subsystems, each featuring dedicated technologies to serve a specific purpose.
During data-taking it is important that data are available from all detector subsystems, and that they are free from integrity issues.
It is possible that data from one or more individual detector components may occasionally be compromised.
This could happen, for example, as a result of a surge in noise from detector electronics or from a high-voltage trip in a detector module.
In order to ensure that data provided for physics analysis are of sufficient quality, any such issues must be identified,
with data collected during the affected time period being marked appropriately.
Thus, data collected during these periods of transient detector issues can be excluded from analyses.
To achieve this, the data collected by each detector subsystem are constantly monitored and their conditions recorded, with automated procedures in place where appropriate.
Communication with detector operation, software, and hardware experts is important to provide a constant feedback loop,
so that the capacity for effectively identifying, preventing or treating data quality (DQ) issues is continuously improved.
The final decision on whether or not  data are usable is based on scrutiny of the relevant issues by a team of experts.

ATLAS began collecting $pp$ collision data at $\sqrt{s}=13$~\TeV\ in mid-2015,
marking the start of the LHC Run~2 data-taking campaign.
During the following four-year experimental programme that ran until the end of 2018, the DQ monitoring procedures, built on the tools and experience developed during Run~1,
were constantly revised and improved.
This, combined with advances in detector operational procedures, resulted in the DQ efficiency increasing year by year.
This paper presents the DQ assessment procedures in place at ATLAS during Run~2, and the DQ performance achieved as a result.
The paper is organised as follows. The ATLAS detector is described in Section~\ref{sec:detector},
and an overview of DQ operations and infrastructure is given in Section~\ref{sec:infra-ops}, where DQ-related tools and applications are introduced.
Following this, the usage of these tools in the context of DQ monitoring and assessment is given in Section~\ref{sec:dqmon}.
Here the DQ workflow, as performed during Run~2 data-taking, is described.
The performance of these practices and resulting DQ efficiencies are presented and discussed in Section~\ref{sec:perf},
with a closing summary provided in Section~\ref{sec:conclusion}.

\section{ATLAS detector}
\label{sec:detector}
 
 
The ATLAS detector at the LHC covers nearly the entire solid angle around the collision point.\footnote{ATLAS uses a right-handed coordinate system with its origin at the nominal interaction point (IP) in the centre of the detector and the $z$-axis along the beam pipe.
The $x$-axis points from the IP to the centre of the LHC ring, and the $y$-axis points upwards.
Cylindrical coordinates $(r,\phi)$ are used in the transverse plane, $\phi$ being the azimuthal angle around the $z$-axis.
The pseudorapidity is defined in terms of the polar angle $\theta$ as $\eta = -\mathrm{ln} \tan(\theta/2)$.
}
It consists of an inner tracking detector surrounded by a thin superconducting solenoid, electromagnetic and hadronic calorimeters,
and a muon spectrometer incorporating three large superconducting toroidal magnets.
 
The inner detector (ID) system is immersed in a \SI{2}{\tesla} axial magnetic field
and provides charged-particle tracking in the range \(|\eta| < 2.5\). The high-granularity silicon pixel detector covers the vertex region and typically provides four measurements per track,
the first hit being normally in the insertable B-layer (IBL), installed before Run~2~\cite{ATLAS-TDR-2010-19,PIX-2018-001}.
It is followed by the semiconductor tracking detector (SCT), which is based on silicon microstrips and usually provides eight measurements per track.
These subsystems are crucial for the reconstruction of charged-particle tracks,
the accuracy of which is limited by the finite resolution of the detector elements and incomplete
knowledge of their positions.
The silicon detectors are complemented by the transition radiation tracker (TRT),
which enables radially extended track reconstruction up to \(|\eta| = 2.0\).
The TRT also provides electron identification information
based on the number of hits (typically 30 in total) above an energy-deposit threshold corresponding to transition radiation.
The ID system provides essential information for the reconstruction of \textit{physics objects} such as electrons~\cite{EGAM-2018-01},
muons~\cite{PERF-2015-10}, $\tau$-leptons~\cite{ATLAS-CONF-2017-029}, photons~\cite{EGAM-2018-01} and jets~\cite{ATL-PHYS-PUB-2015-036},
as well as for identification of jets containing $b$-hadrons~\cite{FTAG-2018-01},
and for event-level quantities that use charged-particle tracks as input.
 
The calorimeter system covers the pseudorapidity range \(|\eta| < 4.9\).
Within the region \(|\eta|< 3.2\), electromagnetic calorimetry is provided by barrel and
endcap high-granularity lead/liquid-argon (LAr) calorimeters,
with an additional thin LAr presampler covering \(|\eta| < 1.8\),
to correct for energy loss in material upstream of the calorimeters.
Hadronic calorimetry is provided by the steel/scintillator-tile calorimeter,
segmented into three barrel structures within \(|\eta| < 1.7\), and two copper/LAr hadronic endcap calorimeters.
The solid angle coverage is completed with forward copper/LAr and tungsten/LAr calorimeter modules
optimised for electromagnetic and hadronic measurements, respectively.
The calorimeters play an important role in the reconstruction of physics objects such as photons, electrons, $\tau$-leptons and jets,
as well as event-level quantities such as missing transverse momentum~\cite{ATLAS-CONF-2018-023}.
 
The muon spectrometer (MS) comprises separate trigger and
high-precision tracking chambers measuring the deflection of muons in a magnetic field generated by superconducting air-core toroids.
The field integral of the toroids ranges between \num{2.0} and \SI{6.0}{\tesla\metre}
across most of the detector.
A set of precision chambers covers the region \(|\eta| < 2.7\) with three layers of monitored drift tubes (MDTs),
complemented by cathode strip chambers (CSCs) in the forward region, where the background is highest.
The muon trigger system covers the range \(|\eta| < 2.4\) with resistive-plate chambers (RPCs) in the barrel, and thin-gap chambers (TGCs) in the endcap regions.
 
Interesting events are selected to be recorded by the first-level (L1) trigger system implemented in custom hardware,
followed by selections made by algorithms implemented in software in the high-level trigger (HLT)~\cite{TRIG-2016-01}.
The L1 trigger makes decisions at the \SI{40}{\MHz} bunch crossing rate and accepts events at a rate below \SI{100}{\kHz},
which the HLT further reduces in order to record complete physics events to disk at about \SI{1}{\kHz}.
The trigger system makes use of physics object- or event-level selections using information from the detector subsystems; hence issues affecting detector components can impact the trigger performance downstream.
Comprehensive monitoring of trigger rates, detector status and physics object quality is therefore imperative to enable the identification of operational issues in real time.
 
The reconstruction of physics objects requires input from a combination of detector subsystems.
To monitor physics object quality, expertise is required in both the combination of detector information used in reconstruction, and the algorithms that facilitate this.
\textit{Combined performance} experts work in collaboration with dedicated subsystem and trigger experts to ensure scrutiny at every level.


\section{Data quality infrastructure and operations}
\label{sec:infra-ops}
\begin{figure}[htb]
\begin{center}
\includegraphics[width=0.8\linewidth]{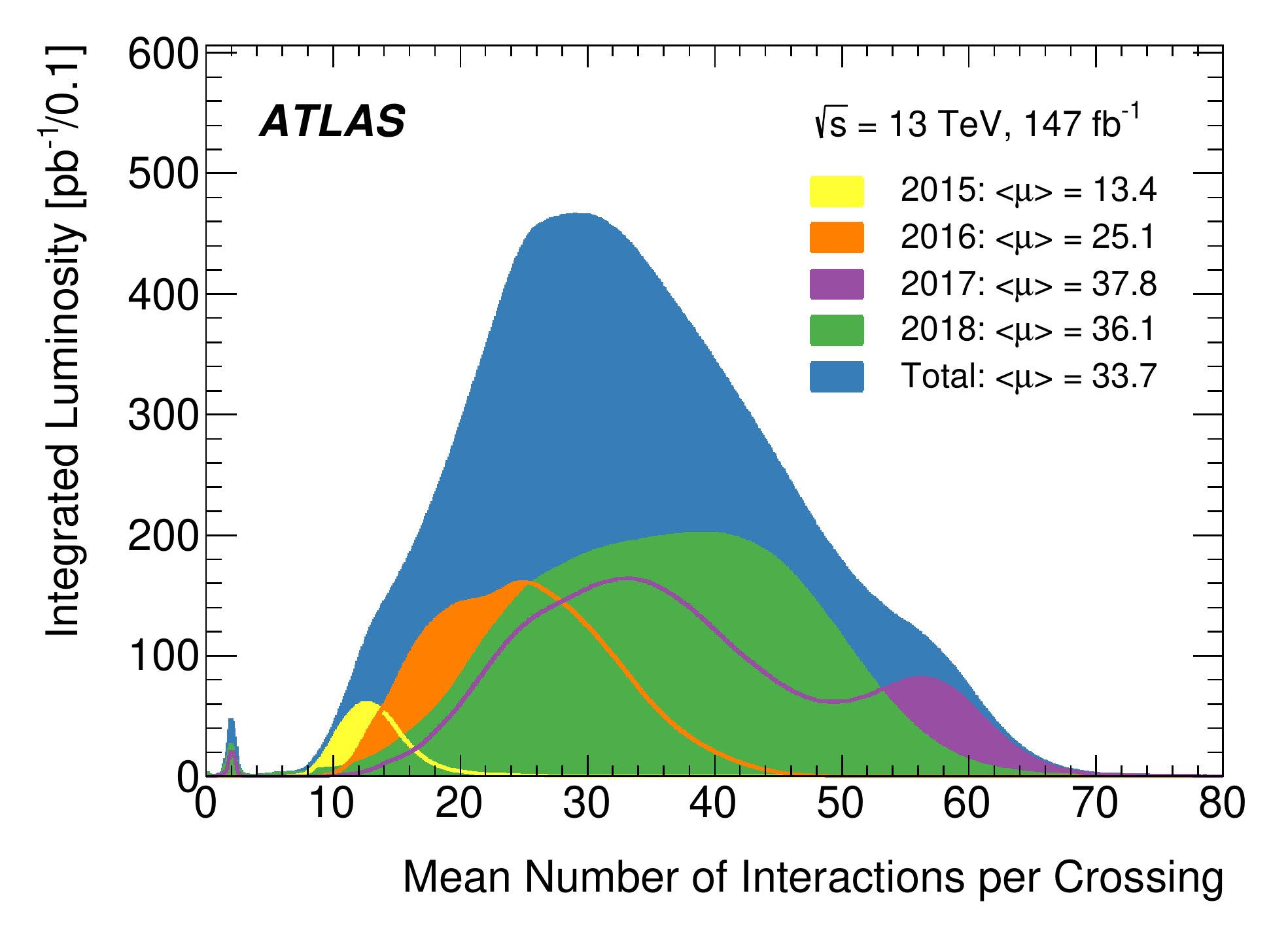}
\caption{Luminosity-weighted distribution of the mean number of interactions per bunch crossing, $\mu$, for the full Run~2 $pp$ collision dataset at $\sqrt{s}=13$~\TeV.
The $\mu$ corresponds to the mean of the Poisson distribution of the number of interactions per crossing calculated for each proton bunch.
It is calculated from the instantaneous per bunch luminosity.
All data recorded by ATLAS during stable beams are shown, including machine commissioning periods, special runs for detector calibration,
LHC fills with a low number of circulating bunches or bunch spacing greater than 25~ns. The integrated luminosity and the mean $\mu$ value for each year are given also.}
\label{fig:mu}
\end{center}
\end{figure}

The ATLAS experiment began taking 13~\TeV\ $pp$ collision data with a nominal bunch-spacing of 50~ns in July 2015,
before the LHC progressed to the design 25~ns bunch-spacing one month later.
The latter is the main data-taking mode during Run~2, referred to hereafter as the ``standard'' Run~2 configuration (in which the ``standard'' Run~2 dataset was collected).
In addition to this, the four years were interspersed with special data-taking periods,
with the LHC providing collision data in different machine configurations.
Such periods include, for example, instances where the LHC collides heavy ions, periods in which the proton centre-of-mass energy is lower than the nominal 13~\TeV,
or periods during which the mean number of interactions per bunch crossing,\footnote{The mean number of interactions per crossing corresponds to the mean of the Poisson distribution of the number of interactions per crossing calculated for each proton bunch.
It is calculated from the instantaneous luminosity per bunch as $\mu=L_{\mathrm{bunch}} \times \sigma_{\mathrm{inel}} / f_{\mathrm{r}}$,
where $L_{\mathrm{bunch}}$ is the per bunch instantaneous luminosity, $\sigma_{\mathrm{inel}}$ is the inelastic cross section (which is taken to be 80~mb for 13~\TeV\ collisions),
and $f_{\mathrm{r}} = 11245.5$~Hz is the LHC revolution frequency.}
$\mu$, is lowered to $\mu=2$.
The luminosity-weighted $\mu$ distribution for the full Run~2 $pp$ collision $\sqrt{s}=13$~\TeV\ dataset is shown in Figure~\ref{fig:mu},
where the $\mu=2$ periods can be seen at the left end of the figure.
A summary of the main data-taking campaigns for 2015--2018 considered in this paper is presented in Table~\ref{tab:considered-periods}.
Other special data-taking periods and configurations that are not discussed herein include machine commissioning periods,
detector calibration or van der Meer (vdM) beam separation scans for luminosity calibration~\cite{ATLAS-CONF-2019-021},
very-low-$\mu$ data-taking for forward physics programmes, and data-taking with high $\beta^{*}$~\cite{AccPhys}.
During 2018, short vdM-like scans (called \textit{emittance scans}) were performed during standard physics data-taking at ATLAS in addition to dedicated vdM scans.
Since emittance scans take place during standard $pp$ data-taking, they are included as a part of the standard Run~2 dataset in this paper,
although data collected during an emittance scan are not used for typical physics analyses (as discussed further in Section~\ref{sec:perf}).
The uncertainty in the integrated luminosity of the standard Run~2 dataset is 1.7\%~\cite{ATLAS-CONF-2019-021},
obtained using the LUCID-2 detector~\cite{LUCID2} for the primary luminosity measurements.

\begin{table}[h]
\small
\begin{center}
\caption{Summary of the main data-taking campaigns of each year during Run 2, along with the corresponding delivered and recorded integrated luminosities.
The integrated luminosity of the data delivered to ATLAS by the LHC between the time at which the LHC has declared ``stable beams'' and the point when sensitive subsystems are switched off to allow a beam dump or beam studies is typically defined as the ``delivered'' luminosity.
The subset of these data that is recorded by ATLAS is defined as the ``recorded'' luminosity.
The corresponding centre-of-mass energy is also given alongside each dataset, where $\sqrt{s_{NN}}$ is the centre-of-mass energy per nucleon pair in the case of heavy-ion data-taking periods.}
\vspace{0.25cm}
\begin{tabular}{|l|l|r@{\hspace{-1ex}}l|r@{\hspace{-1ex}}l|}
\hline
Year 	&Dataset 					       & \multicolumn{4}{c|}{Integrated Luminosity} \\ \cline{3-6}
&  						       & \multicolumn{2}{c|}{Delivered} & \multicolumn{2}{c|}{Recorded}\\ \hline\hline
2015	& $pp$ @ $\sqrt{s}=13$~\TeV\ (50~ns)                   & \tablenum{102.2} &~pb$^{-1}$   & \tablenum{94.5} & ~pb$^{-1}$  \\ 
&$pp$ @ $\sqrt{s}=13$~\TeV\ (25~ns)	 		       & \tablenum{3.88} & ~fb$^{-1}$   & \tablenum{3.63} & ~fb$^{-1}$   \\ 
&$pp$ @ $\sqrt{s}=5.02$~\TeV\	                               & \tablenum{26.1} & ~pb$^{-1}$   & \tablenum{25.6} & ~pb$^{-1}$ \\ 
& Pb-Pb	@ $\sqrt{s_{NN}}=5.02$~\TeV\                           & \tablenum{0.51} & ~nb$^{-1}$   & \tablenum{0.50} & ~nb$^{-1}$  \\
\hline\hline
2016 & $pp$ @ $\sqrt{s}=13$~\TeV\	                       & \tablenum{38.0} & ~fb$^{-1}$   & \tablenum{35.5} & ~fb$^{-1}$ \\ 
& $p$-Pb @ $\sqrt{s_{NN}}=8.16$~\TeV\	                       & \tablenum{170}  & ~nb$^{-1}$   &  \tablenum{167} & ~nb$^{-1}$  \\ 
& $p$-Pb @ $\sqrt{s_{NN}}=5.02$~\TeV\	                       & \tablenum{0.44} & ~nb$^{-1}$ &   \tablenum{0.43} & ~nb$^{-1}$  \\ 
\hline\hline
2017 & $pp$ @ $\sqrt{s}=13$~\TeV\	                       & \tablenum{49.0} & ~fb$^{-1}$ &  \tablenum{46.4} & ~fb$^{-1}$  \\  
& Xe-Xe @ $\sqrt{s_{NN}}=5.44$~\TeV\                           & \tablenum{1.97} & ~nb$^{-1}$ &  \tablenum{1.96} & ~nb$^{-1}$  \\
& $pp$ @ $\sqrt{s}=5.02$~\TeV\ ($\mu=2$)                       & \tablenum{273}  & ~pb$^{-1}$ &  \tablenum{270}  & ~pb$^{-1}$   \\  
& $pp$ @ $\sqrt{s}=13$~\TeV\ ($\mu=2$)                         & \tablenum{150}  & ~pb$^{-1}$ &  \tablenum{148}  & ~pb$^{-1}$  \\  
\hline\hline
2018 &$pp$ @ $\sqrt{s}=13$~\TeV\	 		       & \tablenum{62.1} & ~fb$^{-1}$ &  \tablenum{60.0} & ~fb$^{-1}$  \\  
&$pp$ @ $\sqrt{s}=13$~\TeV\ ($\mu=2$) 	                       & \tablenum{213}  & ~pb$^{-1}$ &  \tablenum{208}  & ~pb$^{-1}$	\\  
& Pb-Pb	@ $\sqrt{s_{NN}}=5.02$~\TeV\                           & \tablenum{1.78} & ~nb$^{-1}$ &  \tablenum{1.74} & ~nb$^{-1}$  \\  
\hline
\end{tabular}
\label{tab:considered-periods}
\end{center}
\end{table}
 
A single uninterrupted period with beams circulating in the LHC machine is called a \textit{fill}.
Typically ATLAS records data continuously during a single LHC fill.
Data-taking for physics begins as soon as possible after the LHC declares ``stable beams'',\footnote{The timestamp that marks the start of ``stable beams'' is determined from the LHC General Machine Timing hardware signal.}
a condition indicating that stable particle collisions have been achieved.
At this point there is high-voltage (HV) ramp up in the Pixel, SCT, and muon system.
Once the preamplifiers of the pixel system are turned on, ATLAS is declared ``ready for physics''.
Each of the datasets taken while ATLAS is continuously recording is referred to as an ATLAS \textit{run}, with each individual run being assigned a unique six-digit run number.
Each run is further divided into \textit{luminosity blocks} (LBs).
An LB is a period of time during which instantaneous luminosity, detector and trigger configuration and data quality conditions are considered constant.
In general one LB corresponds to a time period of $60$~s,
although LB duration is flexible and actions that might alter the run configuration or detector conditions trigger the start of a new LB before $60$~s have elapsed.
LB start and end timestamps are assigned in real time during data-taking by the ATLAS Central Trigger Processor~\cite{Bertelsen:2016yag}.
The LB serves as a standard unit of granularity for data quality book-keeping.
Despite this, rejection of inferior-quality data is possible at a much finer granularity (either event-by-event or within $\mathcal{O}(\mathrm{ms})$ ``time windows'') provided that the grounds and associated parameters are identified in time to be taken into account during the data reconstruction,
as described further in Section~\ref{sec:dqmon}.
 
Detector and trigger status, configuration, and other time-dependent information such as calibration constants, called detector \textit{conditions}~\cite{conditions},
are stored in the ATLAS conditions database~\cite{condDB}.
This is an Oracle database hosting a COOL technology schema~\cite{Trentadue_2012}, which allows the storing of conditions data according to an \textit{interval of validity} (IOV).
The IOV has a start and end timestamp (in~ns), or run and luminosity block identifier, between which the stored conditions are valid and applicable to the data.
Given that LB length is not fixed, the timestamps that mark the start and end of each LB are also stored in the conditions database.
 
A number of online applications are responsible for recording the status of conditions in the conditions database.
Here ``online'' refers to services that run during real-time data-taking at ATLAS,
while ``offline'' applications are run during data reconstruction at Tier~0.\footnote{ATLAS uses the Worldwide LHC Computing Grid (WLCG)~\cite{Bird:2005js} to store and reconstruct data.
The WLCG defines a hierarchical ``Tier'' structure for computing infrastructure: ATLAS uses the Tier~0 farm at CERN for initial (prompt) reconstruction of the recorded data;
later data reprocessing proceeds primarily on the eleven Tier~1 sites.}
The operational status of detector hardware components is controlled and monitored by the Detector Control System (DCS)~\cite{DCS},
providing information that includes component temperatures, high-voltage values and operational capacity.
Further monitoring of low-level detector information is provided by GNAM~\cite{GNAM,Ref-GNAM},
a lightweight tool that monitors detector status at various stages of the data-flow~\cite{Ref-emon}, creating monitoring histograms.
During data-taking the trigger system also constantly monitors the data passed by the L1 trigger,
including data that are not subsequently recorded to disk.
This ensures that interesting events are not unintentionally discarded, and that the HLT algorithms are operating as expected.
In addition, the HLT system generates a set of histograms that monitor aspects of its operation.
For monitoring of higher-level information, including the quality of reconstructed physics objects,
full event reconstruction is run on a small subset of trigger-accepted events and histograms are produced with code similar to that used for monitoring during the offline processing.
Automated checks are carried out on the produced histograms using the Data Quality Monitoring Framework (DQMF)~\cite{DQMF}.
These checks are described in detail in Section~\ref{sec:online}.
The algorithms that run in the DQMF must be lightweight such that they comply with data-processing time restrictions and must be written and tested offline,
independently of online data-taking.
This means that the DQMF is compatible with the online and offline environments,
and its use in the two environments differs only in the methods used to deliver histograms to the framework and publish the results.
 
The Information Service (IS) is used to retrieve monitoring data from the Trigger and Data Acquisition (TDAQ)~\cite{TDAQ} system
and temporarily stores monitoring data in the form of status flags and counters to be shared between online applications.
Histograms created by GNAM, as well as the monitoring histograms and results from the DQMF,
are also sent to IS and can be retrieved~\cite{Ref-OH,Ref-Gatherer} to be rendered by dedicated monitoring display tools.
The monitoring display tools provide an interface for the crew of \textit{shifters}, who are at work in the ATLAS control room 24 hours a day,\footnote{Each 24 hour period is split into three control room shift blocks with an eight hour rotation.}
to monitor data-taking in real time.
Should these online systems indicate an issue with the trigger or a particular detector subsystem,
experts are on call to resolve problems that concern their system to minimise the impact on the quality of ATLAS data.
 
The two major monitoring display tools that are used online in the control room are the Data Quality Monitoring Display (DQMD)~\cite{DQMD} and the Online Histogram Presenter (OHP)~\cite{OHP}.
The DQMD uses the results from the DQMF and presents them in a hierarchical manner, along with the corresponding monitoring histograms and details of the associated DQMF checks.
This provides a global view of the detector status, with coloured status flags indicating the quality of the data according to the DQMF results.
The OHP is a highly configurable histogram presenter, which can be used to display a large number of histograms to be monitored by subsystem, trigger, or DQ experts.
The OHP permits the display of any published histogram in an arbitrary manner, while DQMD strictly visualises the histograms used by the DQMF in a hierarchical structure.
All monitoring information is archived~\cite{Ref-MDA} such that status flags and other information can be propagated offline for DQ documentation.

ATLAS data that pass trigger requirements are organised into \textit{streams},
where a stream is defined as a set of trigger selections and prescales and contains all events that have been stored to disk after satisfying any of those selections.
This streaming is inclusive,\footnote{The exception to this inclusive streaming is the \textit{Debug} stream, which contains events that have passed the L1 trigger, but encounter errors at HLT or DAQ level. This ensures that data encountering problems in the TDAQ system can be recovered. Events collected in this stream are not available in any other stream. Events entering the Debug stream constituted a fraction of less than $10^{-7}$ of the recorded dataset between 2015 and 2018, though physics analyses are asked to consider this stream to avoid missing potentially interesting events.} such that a single event can be included in multiple streams should it satisfy the corresponding selection criteria.
Some streams exist for specific calibration purposes (e.g.\ electronic noise monitoring~\cite{LARG-2013-01}, beam-spot position~\cite{PERF-2015-01}, detector alignment~\cite{ATL-PHYS-PUB-2015-031}),
while others are dedicated to collecting potentially interesting data for physics analysis.
The \textit{Express} stream contains a representative subset of the data collected in the physics and calibration streams,\footnote{The \textit{Express} stream contains a minimal rate ($\sim$$20$~Hz) of events as required to perform a reliable first-validation of the detector subsystems and reconstructed physics objects.}
and is promptly reconstructed at the ATLAS Tier 0 computing facility alongside most calibration streams.
As reconstruction of these streams begins in parallel with data-taking, the data can be made available for offline DQ validation as quickly as possible,
with monitoring histograms produced during the reconstruction process subsequently being analysed by the offline DQMF.

\begin{figure}[htb!]
\begin{center}
\includegraphics[width=\linewidth]{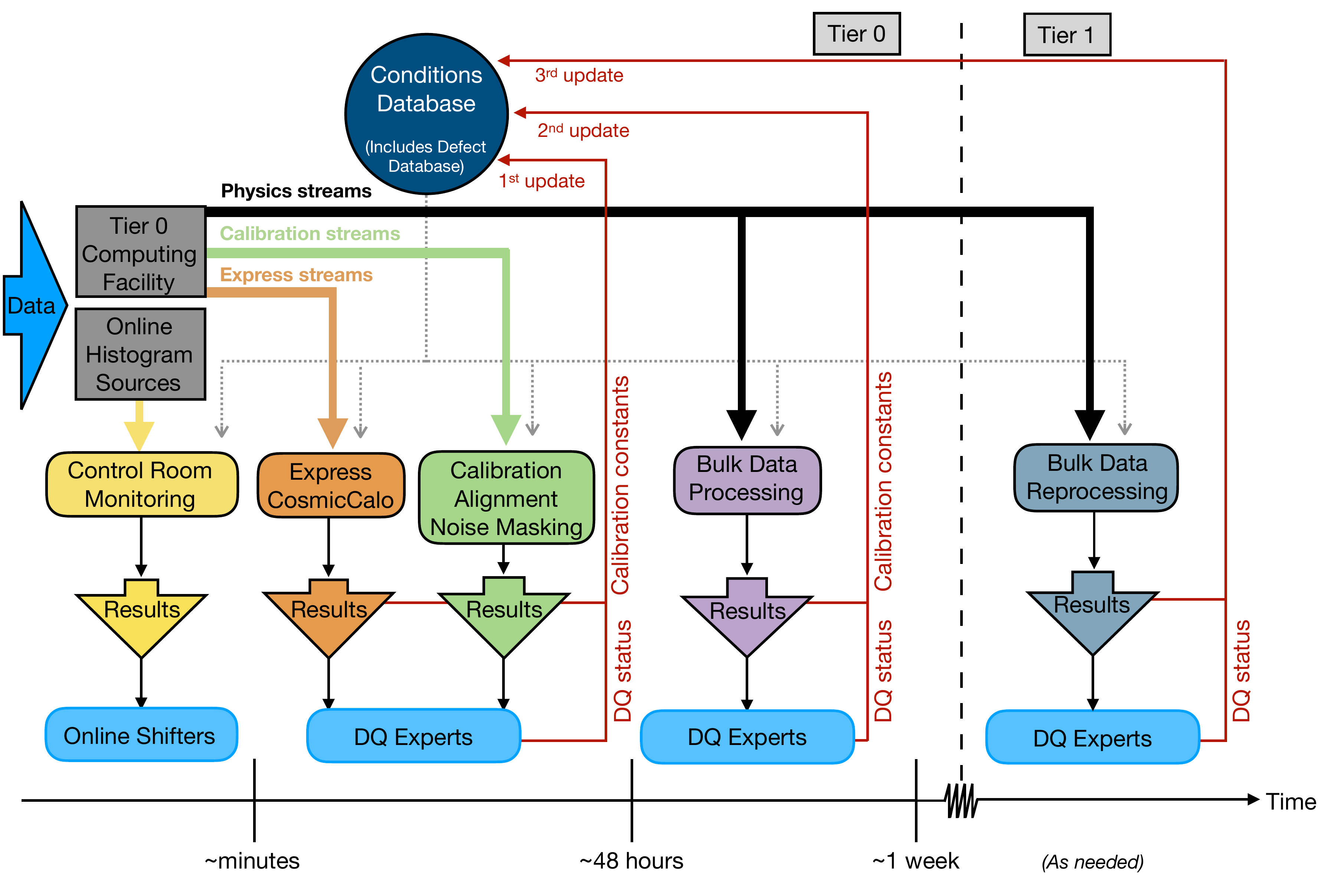}
\caption{Schematic diagram illustrating the nominal Run~2 operations workflow for the data quality assessment of ATLAS data.
Online histogram sources include the high-level trigger farm, the data acquisition system, and full reconstruction of a fraction of events accepted by the trigger.}
\label{fig:dqopscheme}
\end{center}
\end{figure}

The DQ operational workflow is illustrated schematically in Figure~\ref{fig:dqopscheme}.
First, data are monitored online in real time using tools such as DQMD and OHP to display information to monitoring experts (``online shifters'' in Figure~\ref{fig:dqopscheme}) and alert them to the presence of issues.
This first stage seeks to minimise data losses during data-taking, and is discussed in more detail in Section~\ref{sec:online}.
Once the data have been recorded and reconstructed, a two-stage offline DQ assessment ensues.
The first-pass assessment makes use of the promptly reconstructed subset of data from the Express and calibration streams (as illustrated by the orange and green boxes in Figure~\ref{fig:dqopscheme}).
 
Based on this first look at the monitoring output of the data reconstructed offline,
updates can be made to the conditions database by system DQ experts (as indicated by the arrows pointing to the conditions database in Figure~\ref{fig:dqopscheme}),
to be taken into account in the next stage of data processing.
The implementation of these updates, and the procedures followed to ensure their validity, are described in detail in Section~\ref{sec:assessment}.
Upon completion of this first iteration of checks, all streams, including the physics streams, are reconstructed (as shown in the ``bulk data processing'' box in Figure~\ref{fig:dqopscheme}),
making use of the new information gleaned from the first-pass DQ assessment.
Once this full processing is complete a thorough assessment is made using the full data statistics available in the physics streams.
The outcome of this iteration of DQ checks forms the basis for the decision as to which data are usable for physics analysis, and which must be discarded.
A detailed walkthrough of this workflow is provided in Section~\ref{sec:dqmon},
where both online monitoring (Section~\ref{sec:online}) and offline DQ assessment (Section~\ref{sec:assessment}) are discussed.
 
The task of assessing the monitoring data associated with all detector components, and subsequent reconstructed objects, is shared among detector, trigger, and combined performance specialists,
who constitute the ``DQ Experts'' indicated in Figure~\ref{fig:dqopscheme}.
While the detector specialists provide system-specific expertise, the combined performance assessment is split between several subgroups.
The ID Global group validates the combined performance of the ID subsystems and track reconstruction.
A combined calorimeter objects group provides monitoring of reconstructed electrons, photons, $\tau$-leptons, jets, and missing transverse momentum.
Muon combined performance is monitored alongside the muon subsystems within the muon detector group.
A dedicated \textit{$b$-tagging} group is responsible for overseeing the DQ assessment of jets identified as containing $b$-hadrons.
In the case of the trigger group,
DQ assessment responsibilities are divided internally among several combined performance subgroups.
Each of these is responsible for monitoring triggers that select specific physics objects or event-level quantities.
The internal structure of the trigger group is similar to that of the central combined performance monitoring groups.
The trigger and detector subsystems have dedicated shifters performing online monitoring in the ATLAS control room, in addition to the central online DQ shifter.
For example, the ID shifter, responsible for monitoring the ID subsystems, has experience with the ID-specific hardware and software, and can provide more expert input should an issue arise.
In the case of such an issue, all trigger and detector subsystems have dedicated on-call experts available to aid the control room shifters.
Online monitoring of combined performance objects is, for the most part, performed by the online DQ shifter.
There are exceptions to this.
For example,
owing to the large overlap between muon subsystem and muon reconstruction monitoring,
the online muon shifter is responsible for the surveillance of muon object performance in addition to detector monitoring.
 

\FloatBarrier
\section{Data quality monitoring and assessment}
\label{sec:dqmon}
The final product resulting from the DQ monitoring and assessment of ATLAS data is the so-called \textit{Good Runs List} (GRL),
a set of XML~\cite{XML} files that contain the list of LBs that are certified for use in physics analyses for given runs taken over a given time period.
These files are used to filter out data compromised by anomalous conditions, and are fully integrated into the analysis tools used by the ATLAS Collaboration.
The integrated luminosity of a physics dataset, the so-called ``good for physics'' integrated luminosity, is calculated from the LBs that are included in these files.
A GRL is created by querying the DQ status of the data at any given time using the ATLAS defect database~\cite{defectDB}.
The mechanism for recording this information in the database is described in Section~\ref{sec:defects},
and the procedures followed to assess the DQ status of the data are detailed in Section~\ref{sec:assessment}.
 
\subsection{Data quality defects}
\label{sec:defects}
Data quality conditions are recorded by the setting of \textit{defects} in the defect database, a subcomponent of the conditions database.
A defect is set to denote an IOV (at LB-level granularity) where detector conditions are not nominal.
It does not necessarily follow that data need be rejected as a consequence of a defect having been set.
Although serious defects will result in data loss, a defect can also be set as an issue-tracking measure to ease data review in the future.
Defects are stored in versioned COOL tags.
The versioning mechanism both ensures the reproducibility of results and provides flexibility for defects to evolve over time.
The latter can happen as DQ-related issues are addressed or the understanding of a detector problem is improved.
 
There are two different types of versioned defects.
The \textit{primary defects} are usually set manually by DQ and subsystem experts as a consequence of the assessment procedures that are described in Section~\ref{sec:assessment}.
In some cases, primary defects are uploaded automatically, based on information from the DCS, for example.
Primary defects are either \textit{present} or \textit{absent} for a given IOV,
with the default state being absent unless the defect is uploaded during the DQ review.
These primary defect values are associated with a specific iteration of the data processing,
such that they are linked to a time period with uniform conditions.
The \textit{virtual defects} define the logic that is used to determine whether or not the presence of a primary defect can be tolerated,
or if it is necessary to reject data from analysis as a consequence of that defect having been set.
Virtual defects are defined by a logical combination of primary defects or other virtual defects.
In contrast to primary defects, virtual defects are not stored in the database, but are computed upon access.
The virtual defect definitions tend to evolve more slowly than the values of the primary defects.
For example, all LAr calorimeter system defects that are serious enough to warrant the rejection of data from physics use are used to define a single LAr calorimeter virtual defect.
If any one of these serious LAr defects is set for a given IOV, the corresponding LAr virtual defect is also set by definition, as detailed in Ref.~\cite{defectDB}.
In this way the virtual defect contains the information referencing which primary defects describe conditions where the corresponding data should not be used.
This simplifies the process of querying the defect database when creating a GRL.

For automatic defect assignments, data from the DCS archive are used as input to a DCS defect calculator program, which interprets
DCS information to assign defects that indicate suboptimal status of the detector hardware.
For example, the values of the ATLAS solenoid and toroid magnet currents are continuously sampled and stored in the DCS archive.
Should these currents be below a nominal threshold a primary defect is automatically uploaded to the defect database by the DCS defect calculator to indicate that the magnets are not at nominal magnetic field.
The corresponding IOV for which this defect is present denotes the period of time for which the magnet currents are found to be below threshold.
Similar checks exist for the detector subsystems, such as the ID and MS, which ramp up only after the LHC has declared stable beam conditions.
In order to establish whether, for example, the RPC subsystem has reached nominal HV status and is therefore ready for physics data,
the number of RPC HV channels that have reached the desired voltage is monitored by the DCS calculator, using information from the DCS archive.
Once this number exceeds a predefined threshold the HV ramp up can be deemed ``complete'' for the RPC subsystem.
Until then, from the point at which the LHC declares stable beams,
the DCS calculator will automatically set a defect to document that data collected during this time period have been taken while the RPCs were still ramping up.
 
In general, a single primary defect is defined so as to be applicable to a given type of issue that can occur during data-taking.
In the case of the RPC subsystem example above, a specific defect exists to document that the RPC system is still in the process of ramping up to become ready for data-taking.
The meaning of the defect and the class of situations to which it is applicable is documented in the defect database for each defect by way of a \textit{description} field.
Metadata accompany each entry made to the defect database.
This includes an entry timestamp, the identification of the person or process that uploaded the defect itself,
and a comment input by the uploader to add auxiliary information as to the particular issue or occurrence that necessitates the defect upload.
In Run~2 there were a total of $\sim$$450$ different primary defects defined that could be used to track DQ issues with ATLAS data.
 
\subsection{Online data quality monitoring}
\label{sec:online}
Data quality monitoring begins in real time in the ATLAS control room.
Online shifters on duty serve as a first line of defence to identify serious detector-related issues.
Should an issue occur that requires an intervention during data-taking it is important that it is identified quickly to minimise data loss.

Monitoring information stored in the IS can be retrieved to allow DQ monitoring at various levels of the ATLAS data-flow.
The subset of Express stream data reconstructed online is quickly made available to the online shifters via the display tools described in Section~\ref{sec:infra-ops}.
The DQMF tests performed on the reconstruction output can include compatibility checks between the observed distributions from the monitoring data and so-called \textit{reference} histograms, which are
corresponding monitoring histograms taken from a past run that is both free of DQ issues and taken with similar machine operating conditions.
Other tests might involve checks on the number of bins in a histogram above a predefined threshold, or checks on the gradient of a distribution.
For example, histograms that monitor read-out errors should always be empty under ``ideal'' conditions.
If a bin in such a histogram has a non-zero number of entries, a flag would be raised to alert the shifter to the problem.
Online event reconstruction also allows control room DQ shifters to monitor reconstructed physics objects, such as electrons or ID tracks,
permitting real-time monitoring of combined performance in addition to detector status.
For example, if the hit occupancy per unit area in the Pixel, SCT and TRT subsystems is uniform, but a hole or ``cold spot'' is visible in histograms that monitor track occupancy,
this could point to a localised calibration issue that may be difficult to spot using individual subsystem monitoring alone.
The DQMF takes the results of tests on individual histograms and propagates them upwards through a tree,
resulting in a set of top-level status flags, which can be viewed on the DQMD, that alert shifters to potential problems in various subdetectors.
Monitoring histograms are updated to include additional data every few minutes as newly available data are reconstructed.
In this way, online monitoring allows hardware- or software-related issues to be caught in real time and rectified so as to minimise the impact on collected data.

\subsection{Data quality assessment}
\label{sec:assessment}
 
Daily notification emails are distributed to offline data quality shifters, containing information about the current processing status of recent ATLAS runs.
This includes details of which runs are ready for DQ assessment.
The subsequent two-stage assessment workflow follows that introduced in Section~\ref{sec:infra-ops} and illustrated in Figure~\ref{fig:dqopscheme}.
 
Once a run has come to an end and the prompt reconstruction of the Express and calibration streams is complete, the offline DQ assessment begins.
The DQMF extracts information from the reconstruction monitoring output in the form of status flags or numerical results.
These monitoring and reference histograms, as well as the results from the DQMF, are transferred to the offline DQ server,
a virtual machine upon which all subsequent offline DQ operational and assessment procedures take place.
 
Once on the DQ server, the results of the DQMF checks are displayed -- including the output histograms on which the DQMF algorithms are run -- on dedicated,
centrally provided web pages that can be accessed outside of the CERN network.
The web pages are used by detector system, trigger, and combined performance experts to access monitoring information.
Experts can navigate the web pages to view the monitoring histograms for their respective area.
In cases where status flags have been assigned as a consequence of the data quality checks performed by the DQMF,
these are shown alongside the various histograms.
These flags operate under a traffic-light system, with red indicating a potential issue, yellow proposing a manual review of the distribution in question,
and green signalling that the respective automatic checks were passed.
Where relevant, the data of the monitored run are overlaid with data from the assigned reference run.
There is also a mechanism to allow experts to view the same monitoring output from different runs side by side for further comparisons or for investigative purposes.
Comparing the data with reference distributions forms an important part of the validation.

Data quality checks on the Express stream data allow for early scrutiny of a subset of the data collected during a given run,
making it possible to act to correct DQ issues that might be mitigated during data reconstruction.
Should this be the case, by updating calibration constants for example, changes to the conditions can be implemented prior to the physics streams being processed.
This constitutes the first update to the conditions database, as illustrated in Figure~\ref{fig:dqopscheme}.
At this stage, conditions updates ensure that data can still be cleaned, or rejected, at a granularity far finer than a luminosity block.
 
Detector experts use the calibration streams to verify the conditions for detector subsystems,
updating calibration constants and detector alignment information in the conditions database as necessary.
The \textit{SCTNoise} calibration stream contains only events in bunch crossings where the bunches are not filled with protons (so-called \textit{empty} bunch crossings).
Negligible collision activity is expected in these bunch crossings.
Data from this stream are used to identify noisy SCT strips.
Having been identified in this stream, which contains a noise-dominated data sample, these strips can then be masked in the software to prevent noise in the SCT system from biasing track reconstruction.\footnote{The masking of an SCT strip results in the strip being ignored in the track reconstruction.}
Events in the \textit{CosmicCalo} stream are selected on the basis of low-threshold calorimeter activity occurring in empty bunch crossings,
providing a sample enriched with electronic noise from the calorimeter~\cite{LARG-2013-01}.
These data are used alongside dedicated calorimeter calibration streams\footnote{The \textit{LArCellsEmpty} stream contains ``partially built'' events from empty bunch-crossings, where only the LAr cells associated with a high-energy deposit are stored. This results in a reduced event size, permitting both the use of lower energy thresholds for the events to be recorded and larger samples to be stored. The \textit{LArNoiseBurst} stream contains noise burst candidate events, which are identified (and their timestamps marked) at the HLT-level. This early identification, a significant improvement relative to pre-2015 operations, facilitates noise burst rejection during these first DQ checks, yielding a cleaner sample in which to study and identify noisy channels. }
to identify noisy channels that must be masked in reconstruction,\footnote{The masking of a LAr channel results in the energy deposited in that channel being ignored, with the corresponding cell energy instead being estimated from the average energy of the eight neighbouring channels within the same calorimeter layer.}
and to diagnose bursts of noise originating in the calorimeter that must be rejected from physics analyses in a procedure detailed in Ref.~\cite{LARG-2013-01}.
In the latter case, the timestamp at which the burst occurs is identified and all data collected within a window of $\mathcal{O}(\mathrm{ms})$ surrounding that timestamp is marked as being within a noise burst IOV.
By uploading these time windows to the conditions database,
the corresponding data can be marked during the imminent reconstruction of the physics streams so that they can be rejected in physics analyses.
This same procedure is used in order to reject small time windows where data from the LAr system are corrupted.
To account for the impact of rejecting these small time windows from the dataset,
these IOVs are removed when calculating the total ``good for physics'' integrated luminosity of the corresponding dataset.
The data rejected during these time windows are consequently accounted for in the total DQ inefficiency.
Approximately $69$~\ipb\ of data were rejected from the standard $\sqrt{s}=13$~\TeV\ dataset as a consequence of rejecting time windows over the course of Run~2.
 
The conditions to be updated during this first-pass assessment of the offline-reconstructed data also include constants pertaining to the alignment of the ID subsystems~\cite{ATL-PHYS-PUB-2015-031},\footnote{The dedicated \textit{IDTracks} stream is used to obtain a precise determination of the positions of the sensitive ID elements and their geometrical distortions relative to the ideal or designed geometry.}
and the precise determination of the ``beam spot'' location~\cite{PERF-2015-01}, the luminosity-weighted centroid of which defines the average collision point.
The LAr, SCT, beam spot, and alignment folders in the conditions database must be updated with the relevant information before the bulk reconstruction of all physics streams is launched.
 
This process of database updates and primary DQ review is known as the \textit{calibration loop}.
The calibration loop typically lasts $\sim$48~hours from the point at which the first-pass data processing begins,
although this can be further shortened to 24~hours during busy periods.
In exceptional cases the calibration loop can be delayed at the request of detector subsystem experts,
effectively putting full reconstruction on hold until the issue prompting the delay is resolved.
Such situations are rare, but can arise due to problematic database uploads,
excessive detector noise requiring further investigation, or reconstruction failures.
 
\begin{figure}[!h]
\centering
{\includegraphics[width=0.49\textwidth]{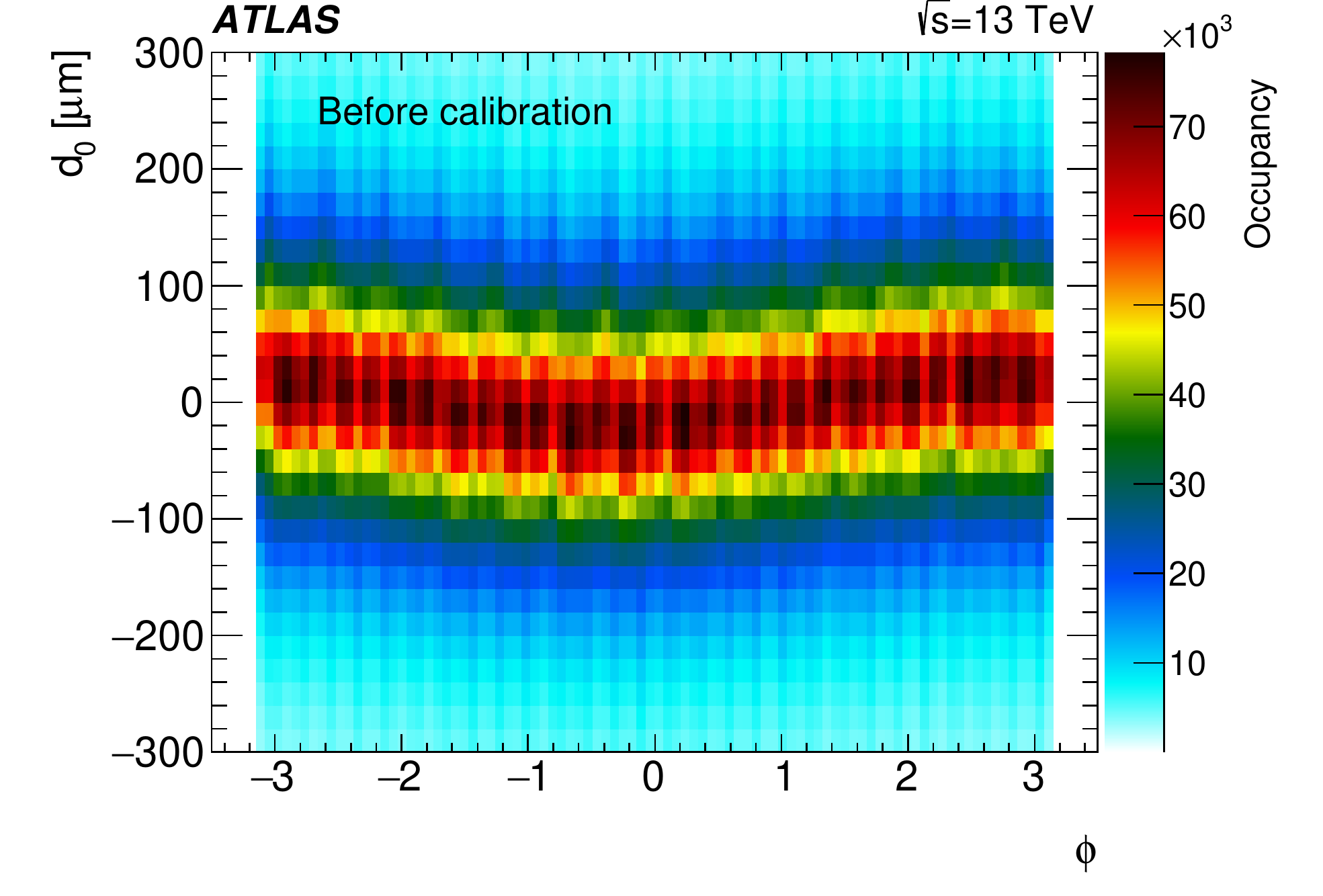}}
{\includegraphics[width=0.49\textwidth]{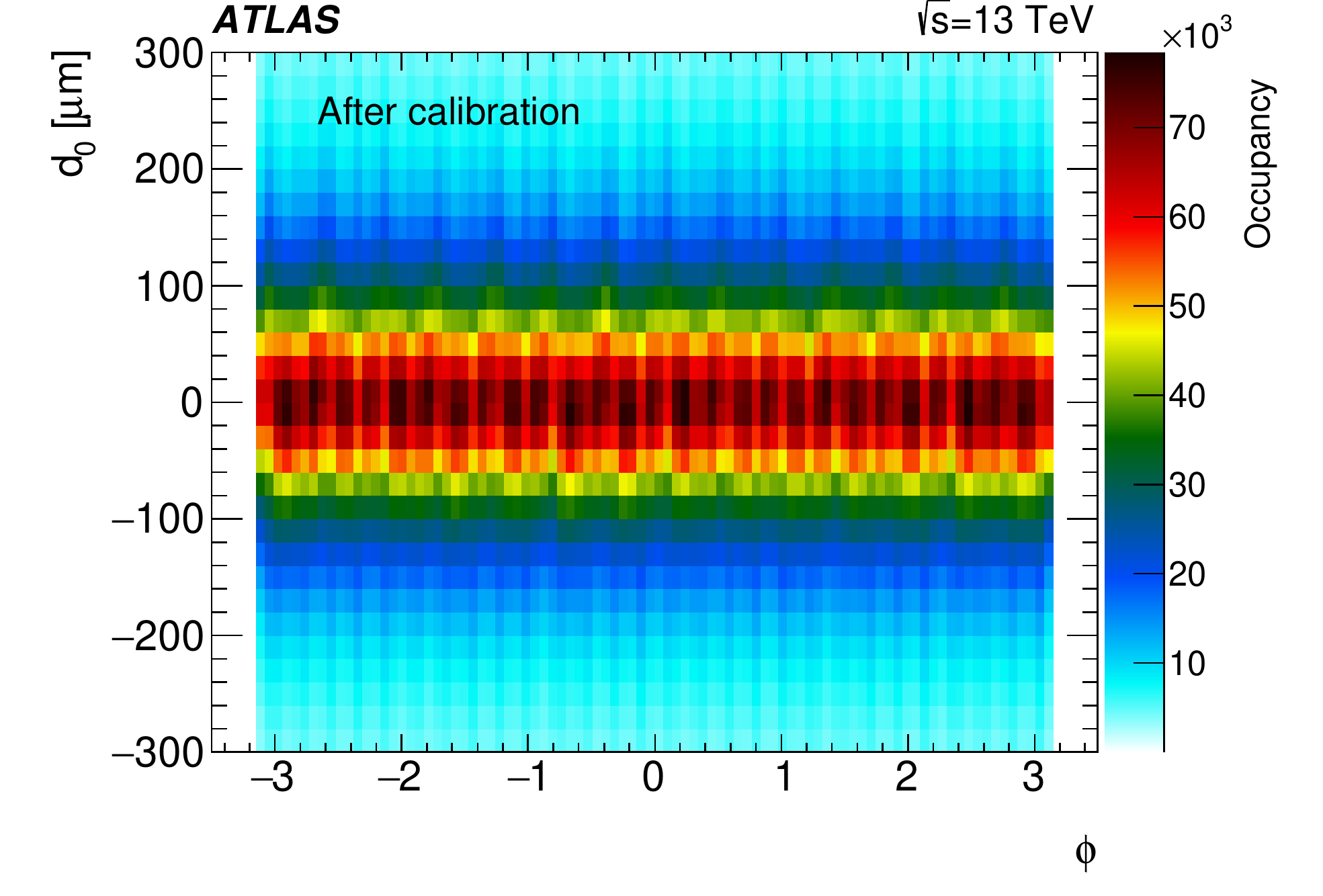}}
\caption{ Transverse impact parameter $d_{0}$ versus $\phi$ for charged-particle tracks. Left: For the first processing of the Express stream,
the beam-spot location is not yet known accurately and $d_0$ is calculated relative to the beam-spot position as determined online. Right: A correctly determined beam spot results in a flat distribution after the application of updated conditions during the calibration loop, since here $d_0$ is calculated relative to the reconstructed beam spot.}
\label{fig:beamspot}
\end{figure}

Once the database conditions are up to date and the calibration loop delay expires,  processing of the physics streams is launched, taking into account these updated conditions.
The reconstructed dataset is usually available after 24--48~hrs,
at which stage a final DQ assessment is performed on the full data statistics.
Detector system and combined performance experts scrutinise the monitoring output once more,
either using the dedicated web-page displays or running stand-alone algorithms directly on the monitoring output.
At this stage it is important not only to validate the quality of the full dataset, but also to ensure that conditions updates during the calibration loop have been applied successfully.
For example, Figure~\ref{fig:beamspot} shows the transverse-plane distance of closest approach of charged-particle tracks to the beamline (the \textit{transverse impact parameter}, $d_0$)
versus $\phi$ before and after the beam-spot determination during the calibration loop.
Following the calibration loop, where updated beam-spot conditions are applied, the distribution should be flat, as illustrated in the right panel of Figure~\ref{fig:beamspot}.

With the full reconstruction complete, any remaining issues that act to the detriment of the DQ at this stage,
such as residual noise or non-negligible losses in coverage, are documented by assigning an appropriate defect to the affected IOV, at luminosity block granularity, in the defect database.

A second update to the conditions database is possible at this stage to provide more complete rejection of LAr noise, for example.
This further update to the conditions would only be applied should the data be subject to a reprocessing, i.e.\ rerunning the reconstruction procedure with an updated configuration.
For the LAr noise example, a conditions update may facilitate a more thorough cleaning of LAr noise using finer-granularity time windows.
In such a case, previously set LAr noise primary defects may be removed from the corresponding LBs as a consequence of this more granular cleaning.
This is an example of a primary defect evolving with a new iteration of the data processing.
The implication here is that data reprocessing campaigns can result in improved DQ efficiency for a given dataset.
Reprocessing campaigns can take place as needed,
but large-scale reprocessings will generally occur once a year at most to incorporate final conditions and any important software improvements.
This brings the reconstruction configuration for a large dataset collected over a long period of time into alignment.
By this stage the data will have been distributed to Tier 1 sites, as indicated on the far right of Figure~\ref{fig:dqopscheme},
and so further data reprocessings do not take place at the Tier 0 computing facility.\footnote{Though data reprocessings are nominally performed at Tier 1 sites to ensure that Tier 0 resources are available for prompt data reconstruction, the Tier 0 can be seamlessly included as an additional ATLAS WLCG site when it is not otherwise in use. }
Any time that the data are reprocessed the DQ validation procedure is repeated to rule out the presence of any insidious effects resulting from the updated conditions or reconstruction software.
Both the output histograms and the results from DQMF are archived for future reference, as is the case for every data processing iteration.
This information, used for the DQ web pages through which a large portion of the offline DQ assessment is performed, is stored in a ROOT~\cite{Brun:1997pa} file format.
These files, for data taken in 2015--2018, including the first-pass processing, bulk data processing, and bulk data reprocessing campaigns, amount to a total of approximately 10~TB of data.
During a single processing of the main physics stream for a single ATLAS run $\mathcal{O}(100$k$)$ monitoring histograms are produced.
While it is important that monitoring information is available in order to quickly study DQ issues,
the number of histograms and their respective granularities are constantly reviewed to minimise the impact on processing time and storage resources.

\begin{table}[h]
\scriptsize
\begin{center}
\caption{Summary of the data quality groups involved in the continuous review and sign-off of ATLAS data.
Included are details pertaining to the nominal shift crew responsible for DQ assessment in each area.
If a dedicated online shifter is responsible for the online monitoring of a system in addition to the online DQ shifter, this is indicated under the ``dedicated control room shifter'' column.
Should a subsystem or group be responsible for routinely updating DQ-related conditions during the calibration loop, this is indicated under ``conditions updates'',
with the corresponding streams monitored for DQ purposes during the calibration loop listed in the neighbouring column.
If any of the defects under the responsibility of a given group are uploaded automatically this is indicated in the last column.}
\resizebox{\textwidth}{!}{
\begin{tabular}{|l|c|c|c|c|c|}
\hline
Group/Subsystem & Dedicated Control  & Offline DQ          &  Conditions Updates       & Calibration Loop    &  Automatic Defect  \\
& Room Shifter       & Shifters \& Experts &  (calibration loop)       & Streams             &  Assignment        \\
\hline\hline
\multirow{2}{*}{Pixel}  & \multirow{6}{*}{ID shifter} & 1 shifter           & \multirow{2}{*}{Disabled module masks}     & \multirow{2}{*}{Express}             & \multirow{2}{*}{Yes}                \\
&                             & 1 expert            &                           &                     &                    \\\cline{1-1}\cline{3-6}
 
\multirow{2}{*}{SCT}    &                             & 1 shifter           & \multirow{2}{*}{Noisy strip masks}         & SCTNoise            & \multirow{2}{*}{Yes}                \\
&                             & 1 expert            &                           & Express             &                    \\\cline{1-1}\cline{3-6}
 
\multirow{2}{*}{TRT}    &                             & 1 shifter           & \multirow{2}{*}{Dead straws}               & \multirow{2}{*}{Express}             & \multirow{2}{*}{No}                 \\
&                             & 2 experts           &                           &                     &                    \\
\hline
\multirow{4}{*}{LAr}    & \multirow{6}{*}{Calorimeter shifter}  &  2 shifters     & Noisy channel masks       & LArCellsEmpty       & \multirow{4}{*}{No}                 \\
&                             & 2 experts            & Noise burst rejection     & LArNoiseBurst       &                    \\
&                             &           & Data corruption           & CosmicCalo          &                    \\
&                             &                     &                           & Express             &                    \\\cline{1-1}\cline{3-6}
 
\multirow{2}{*}{Tile}   &                             & 1 shifter           & \multirow{2}{*}{$-$}      & \multirow{2}{*}{Express}             & \multirow{2}{*}{Yes}                \\
&                             & 1 expert            &                           &                     &                    \\
\hline
\multirow{2}{*}{L1}     & \multirow{4}{*}{Trigger shifter}             & 1 shifter           & \multirow{2}{*}{$-$}      & \multirow{2}{*}{Express}             & \multirow{2}{*}{No}                 \\
&                             & 1 expert            &                           &                     &                    \\\cline{1-1}\cline{3-6}
 
\multirow{2}{*}{HLT}    &                             & 1 shifter           & \multirow{2}{*}{$-$}      & \multirow{2}{*}{Express}             & \multirow{2}{*}{Yes}                \\
&                             & 5 experts           &                           &                     &                    \\
\hline
\multirow{2}{*}{$b$-tagging}         & \multirow{2}{*}{$-$} & 1 shifter           & \multirow{2}{*}{$-$}      & \multirow{2}{*}{Express}             & \multirow{2}{*}{No}                 \\
&                & 2 experts           &                           &                     &                    \\
\hline
\multirow{2}{*}{Calorimeter objects} & \multirow{2}{*}{$-$}  & 1 shifter           & \multirow{2}{*}{$-$}      & \multirow{2}{*}{Express}             & \multirow{2}{*}{No}                 \\
&                & 5 experts           &                           &                     &                    \\
\hline
\multirow{3}{*}{ID Global}  & \multirow{3}{*}{$-$} & 1 shifter           & ID alignment              & IDTracks            & \multirow{3}{*}{Yes}                \\
&                      & 2 experts           & Beam-spot position        & BeamSpot            &                    \\
&                             &                     &                           & Express             &                    \\
\hline
Muons                         & \multirow{2}{*}{Muon shifter}   & 1 shifter           & \multirow{2}{*}{$-$} & \multirow{2}{*}{Express}             & \multirow{2}{*}{Yes}                \\
(subsystem \& performance)    &                & 1 expert            &                 &                     &                    \\
\hline
\multirow{2}{*}{Luminosity}   & \multirow{2}{*}{$-$}  & 1 shifter           & \multirow{2}{*}{$-$} & \multirow{2}{*}{Express}             & \multirow{2}{*}{Yes}                \\
&                & 1 expert            &                 &                     &                    \\\hline
\end{tabular}}
\label{tab:teams}
\end{center}
\end{table}
 
An overview of the DQ monitoring groups and their organisation, introduced in Section~\ref{sec:infra-ops}, is given in Table~\ref{tab:teams}.
Offline DQ assessment typically involves at least one offline shifter and one longer-term expert per subsystem or combined performance area.
As shown in Table~\ref{tab:teams}, in some cases the complexity of the assessment procedure requires larger teams.
For example, the trigger DQ assessment requires input from several physics signature experts;
and the nature and multiplicity of the LAr calibration loop updates requires that the work be shared between two shifters.
In addition to the calibration loop updates already discussed,
both the pixel and TRT subsystems must update the database with information about any newly disabled pixel modules or TRT straws,
such that no activity should be expected in these regions during reconstruction.
Conditions updates from other subsystems are also possible during the calibration loop, although they do not form part of the routine DQ workflow carried out for every ATLAS physics run.
Subsystem and trigger DQ experts and shifters frequently interact with the ATLAS experiment operations teams.
This ensures efficient communication of matters that may influence decisions related to DQ.
Usually it is the DQ experts who take responsibility for uploading defects to the defect database, based on the initial DQ assessment carried out by the offline DQ shifters.
While some subsystems make use of the automatic defect upload facility (by using the DCS calculator for example), combined performance groups do not.
The only exception to this is for beam-spot monitoring;
in cases where an accurate determination of the beam-spot position has not been possible, the corresponding IOV is automatically marked to flag this.
Otherwise, automatic defects primarily mark data affected by detector hardware problems, or periods of time during which a particular detector element is not operational.
 
The full data validation procedure is overseen by ATLAS data quality and data preparation coordination, with formal sign-off of the data taking place on a weekly basis during dedicated meetings.
The validation of the full dataset involves the collaboration of detector, trigger, and physics object reconstruction and performance experts.
Typically, new runs can be fully validated 1--2 weeks after having been delivered by the LHC.

Once a given portion of runs have completed DQ validation (for example, at the close of a data-taking \textit{period})\footnote{Runs are grouped together into \textit{periods} during which data-taking conditions are considered to be uniform.}
a new GRL can be generated and propagated to physics analysis groups so that they can select appropriate data for analysis.
The GRL is generated by querying the defect database for a selection of runs and LBs that satisfy predefined ``good for physics'' conditions.
The ``good for physics'' definition is itself contained in a virtual defect, and so the criteria used to determine whether data enter a GRL are documented in versioned COOL tags.
Safety defects exist to ensure that runs that have not been validated do not enter the GRL.
For example, each run initially has a so-called \texttt{NOTCONSIDERED}\footnote{Data marked with a \texttt{NOTCONSIDERED} defect would not be permissible in physics analyses.} defect assigned for all LBs until it is released to DQ experts for assessment.
For each run taken outside normal data-taking periods, where the data are not meant for physics use in analyses,
the \texttt{NOTCONSIDERED} defect remains in place such that it is not propagated to DQ experts.
Furthermore, each detector system and physics object group that scrutinises the data in the procedure described in this section must manually remove a so-called \texttt{UNCHECKED} defect from the database for each run once their validation is completed.
A GRL cannot be generated while an \texttt{UNCHECKED} defect remains present for any of the data being considered.
 
There are cases where special GRLs are made available to analysers, where the standard ``good for physics'' definition may not be suitable,
or a significant fraction of the recorded data is rejected owing to a defect that can be tolerated by a specific analysis.
For example, physics analyses that do not rely on muon identification may be able to use data taken during periods where the ATLAS toroids were not operational.
In such a case a special GRL can be made that includes data that would otherwise be rejected owing to magnet defects.

 
\FloatBarrier
\section{Data quality performance}
\label{sec:perf}
Detector-, trigger-, reconstruction-, or processing-related problems that result in data being rejected on the grounds of DQ are accounted for in the DQ efficiency.
The DQ efficiency is calculated relative to recorded, rather than delivered, integrated luminosity.
In some cases, detector-related processes or problems result in data not being recorded for physics by ATLAS in the first place.
This includes data taken during the so-called ``warm start'' that the tracking detectors undergo once the LHC declares ``stable beams'',
which involves ramping up the high voltage and turning on the preamplifiers for the pixel system.
The inefficiency due to the time it takes for the pixel preamplifiers to turn on
is included in the ATLAS data-taking efficiency.
This is defined as the fraction of data delivered by the LHC recorded by ATLAS, regardless of the quality of the data.
Should remaining detector subsystems still be undergoing high-voltage ramp up after the warm start is complete,
the inefficiency due to this is accounted for in the DQ efficiency.
 
The data quality efficiency is presented in terms of a luminosity-weighted fraction of good quality data recorded during stable beam periods.
Only periods during which the recorded data were intended to be used for physics analysis are considered for the DQ efficiency.
In particular, this excludes machine commissioning periods, special runs for detector calibration, LHC fills with a low number of circulating bunches or bunch spacing greater than 25~ns,
and periods before beams were separated to achieve low-$\mu$ conditions for low-$\mu$ runs.\footnote{In order to achieve low-$\mu$ conditions the LHC begins colliding the beams head-on,
before slowly separating them in the transverse plane to reduce the average number of interactions per bunch crossing.}
The defect database framework allows defects to be associated with a given detector subsystem, reconstruction algorithm, or operating condition.
DQ defects can be associated with physics objects as well, but in these cases there is generally some underlying detector-level cause behind the issue.
It is therefore possible to express the DQ efficiency during Run 2 data-taking in the context of each detector subsystem,
and to present an overall DQ efficiency for ATLAS during Run 2.

The luminosity-weighted fraction of good quality data delivered during stable beam periods for 2015, 2016,
2017, and 2018 is presented in Tables~\ref{tab:eff-2015},~\ref{tab:eff-2016},~\ref{tab:eff-2017}, and ~\ref{tab:eff-2018}, respectively.
Since the presented efficiencies are weighted according to integrated luminosity the efficiency is dominated by periods of standard $pp$ data-taking at $\sqrt{s}=13$~\TeV.
The DQ efficiencies for data-taking campaigns using other machine configurations, as detailed in Table~\ref{tab:considered-periods} (including $\mu=2$ $pp$ data taken at $\sqrt{s}=13$~\TeV),
are displayed separately in these tables.
The total DQ efficiency improved over the course of Run 2, with 88.8\% in 2015, 93.1\% in 2016,
95.7\% in 2017 and 97.5\% in 2018,
for the (dominant) standard $\sqrt{s}=13$~\TeV\ $pp$ data-taking campaigns of each year.
Combining these efficiencies results in a total data quality efficiency of 95.6\% for the standard Run 2 $\sqrt{s}=13$~\TeV\ $pp$ dataset,
with a corresponding 139~fb$^{-1}$ of data certified as being good for physics analysis.
The DQ-related inefficiencies per subsystem corresponding to this dataset are illustrated in Figure~\ref{fig:run2_eff} and presented in Table~\ref{tab:eff-run2},
and the cumulative ``good for physics'' integrated luminosity versus month in a year is shown alongside the corresponding delivered and recorded integrated luminosities in the top panel of Figure~\ref{fig:intlumivstimeRun2DQall}.
The evolution of the DQ efficiency as a function of collected integrated luminosity is illustrated in the bottom panel of Figure~\ref{fig:intlumivstimeRun2DQall},
where the more significant DQ-loss incidents can be seen as sharp drops in the cumulative DQ efficiency.
For example, the sharp decline from $\sim$$98$\% to $94$\% around the $10$~\ifb\ mark in 2017 is due to three runs where the ATLAS toroids were not operational.
 
Periods during which the toroids are not in operation are typically not used for physics analysis,\footnote{There is an exception to this for analyses that do not rely on muon reconstruction~\cite{HIGG-2016-17,EXOT-2016-19,SUSY-2016-09}.}
although these data are useful for alignment of the muon system, making these periods profitable despite the associated inefficiency indicated in Table~\ref{tab:eff-run2}.
These data can also be used for other detector-related studies or for calibration purposes,
and in these cases the subsystem in question often switches into a non-standard data-taking configuration to facilitate online tests.
The DQ efficiency for the subsystem in question is not impacted as a result of being in a non-standard configuration in these cases,
since these periods occur in the ``shadow'' of a magnet issue and hence are accounted for in the inefficiency quoted for the toroids and/or solenoid.
The same accounting principle is applied in cases where a specific detector subsystem is switched off, for example.
Should other subsystems take advantage of this to perform studies specific to their subsystem, the corresponding DQ inefficiency is assigned only to the subsystem that is off.

\begin{table}[h]
\scriptsize
\begin{center}
\caption{Luminosity-weighted relative detector uptime and good data quality efficiencies (in \%)
during stable beams $pp$ collision physics runs at $\sqrt{s}=13$~\TeV\ and $5.02$~\TeV\ between July and November 2015.
Also shown are 0.49~nb$^{-1}$ of Pb-Pb collision data collected at $\sqrt{s_{NN}}=5.02$~\TeV\ between November and December 2015.
}
\resizebox{\textwidth}{!}{
\begin{tabular}{|l|S[table-format=3.2,table-align-text-post = false,table-number-alignment=center]|S[table-format=3.2,table-align-text-post = false,table-number-alignment=center]|S[table-format=3.2,table-align-text-post = false,table-number-alignment=center]|S[table-format=3.2,table-align-text-post = false,table-number-alignment=center]|S[table-format=3.2,table-align-text-post = false,table-number-alignment=center]|S[table-format=3.2,table-align-text-post = false,table-number-alignment=center]|S[table-format=3.2,table-align-text-post = false,table-number-alignment=center]|S[table-format=3.2,table-align-text-post = false,table-number-alignment=center]|S[table-format=3.2,table-align-text-post = false,table-number-alignment=center]|S[table-format=3.2,table-align-text-post = false,table-number-alignment=center]|S[table-format=3.2,table-align-text-post = false,table-number-alignment=center]|S[table-format=3.2,table-align-text-post = false,table-number-alignment=center]|S[table-format=3.2,table-align-text-post = false,table-number-alignment=center]|}
\hline
\multicolumn{14}{|c|}{\textbf{2015 Data Quality Efficiency [\%]}}  \\ \hline
\multirow{2}{*}{Dataset}&\multicolumn{3}{c|}{Inner Tracker}&\multicolumn{2}{c|}{Calorimeters}&\multicolumn{4}{c|}{Muon Spectrometer}&\multicolumn{2}{c|}{Magnets}&\multicolumn{2}{c|}{Trigger}\\ \cline{2-14}
&\multicolumn{1}{c|}{Pixel} & \multicolumn{1}{c|}{SCT} &\multicolumn{1}{c|}{TRT} & \multicolumn{1}{c|}{LAr} & \multicolumn{1}{c|}{Tile} &\multicolumn{1}{c|}{MDT} & \multicolumn{1}{c|}{RPC}& \multicolumn{1}{c|}{CSC} & \multicolumn{1}{c|}{TGC} &\multicolumn{1}{c|}{Solenoid}& \multicolumn{1}{c|}{Toroid}&\multicolumn{1}{c|}{L1}& \multicolumn{1}{c|}{HLT}\\ \hline\hline
$pp$ @ 13~\TeV\ (50~ns)  &  99.84  &  99.63 & 95.28     &  98.53  &  100  & 95.28  &  100     & 100  &  99.70  & 100 &   95.87  & 100     &  99.94\\
$pp$ @ 13~\TeV           &  93.84  &  99.77 &  98.29  &  99.54  &  100  & 100  &   99.96  & 100  &  99.97  & 100 &   97.79  &  99.97  &  99.76\\ 
$pp$ @ 5.02~\TeV            & 100     & 100    & 100     & 100     &  100  & 100  &   99.96  & 100  &  99.94  & 100 &  100     &  99.24  & 100\\
Pb-Pb @ 5.02~\TeV  &     100  &  100  &  99.64  &   97.57  &  100  &  99.80  &   99.98  &  99.90  &  99.89  &  100  &  100  &  100  &  100\\ \cline{0-13}\noalign{\smallskip}
\cline{5-10}
\multicolumn{1}{c}{}& \multicolumn{3}{c|}{} & \multicolumn{3}{c|}{Data Quality Efficiency [\%]} & \multicolumn{3}{c|}{Integrated Luminosity} \\\cline{0-9}
$pp$ @ 13~\TeV\ (50~ns)  & \multicolumn{3}{c|}{\multirow{4}{*}{Good for Physics}} & \multicolumn{3}{S[table-format=3.2,table-align-text-post = false,table-number-alignment=center]|}{88.77} & \multicolumn{3}{S[table-format=3.2, table-space-text-post={~pb$^{-1}$}] <{~pb$^{-1}$}|}{83.9}  \\
$pp$ @ 13~\TeV\          & \multicolumn{3}{c|}{ } & \multicolumn{3}{S[table-format=3.2,table-align-text-post = false,table-number-alignment=center]|}{88.79} & \multicolumn{3}{S[table-format=3.2, table-space-text-post={~fb$^{-1}$}] <{~fb$^{-1}$}|}{3.22} \\
$pp$ @ 5.02~\TeV\           & \multicolumn{3}{c|}{ } & \multicolumn{3}{S[table-format=3.2,table-align-text-post = false,table-number-alignment=center]|}{99.14} & \multicolumn{3}{S[table-format=3.2, table-space-text-post={~pb$^{-1}$}] <{~pb$^{-1}$}|}{25.3} \\
Pb-Pb @ 5.02~\TeV\           & \multicolumn{3}{c|}{ } & \multicolumn{3}{S[table-format=3.2,table-align-text-post = false,table-number-alignment=center]|}{96.76} & \multicolumn{3}{S[table-format=3.2, table-space-text-post={~nb$^{-1}$}] <{~nb$^{-1}$}|}{0.49} \\  \cline{0-9}
\end{tabular}}
\label{tab:eff-2015}
\end{center}
\end{table}

\begin{table}[h]
\scriptsize
\begin{center}
\caption{Luminosity-weighted relative detector uptime and good data quality efficiencies (in \%)
during stable beams in $pp$ collision physics runs at $\sqrt{s}=13$~\TeV\ between April and October 2016.
Also shown is 165~nb$^{-1}$ of $p$-Pb collision data collected at $\sqrt{s_{NN}}=5.02$ and $8.16$~\TeV\ between November and December 2016.
}
\resizebox{\textwidth}{!}{
\begin{tabular}{|l|S[table-format=3.2,table-align-text-post = false,table-number-alignment=center]|S[table-format=3.2,table-align-text-post = false,table-number-alignment=center]|S[table-format=3.2,table-align-text-post = false,table-number-alignment=center]|S[table-format=3.2,table-align-text-post = false,table-number-alignment=center]|S[table-format=3.2,table-align-text-post = false,table-number-alignment=center]|S[table-format=3.2,table-align-text-post = false,table-number-alignment=center]|S[table-format=3.2,table-align-text-post = false,table-number-alignment=center]|S[table-format=3.2,table-align-text-post = false,table-number-alignment=center]|S[table-format=3.2,table-align-text-post = false,table-number-alignment=center]|S[table-format=3.2,table-align-text-post = false,table-number-alignment=center]|S[table-format=3.2,table-align-text-post = false,table-number-alignment=center]|S[table-format=3.2,table-align-text-post = false,table-number-alignment=center]|S[table-format=3.2,table-align-text-post = false,table-number-alignment=center]|}
\hline
\multicolumn{14}{|c|}{\textbf{2016 Data Quality Efficiency [\%]}}  \\ \hline
\multirow{2}{*}{Dataset}&\multicolumn{3}{c|}{Inner Tracker}&\multicolumn{2}{c|}{Calorimeters}&\multicolumn{4}{c|}{Muon Spectrometer}&\multicolumn{2}{c|}{Magnets}&\multicolumn{2}{c|}{Trigger}\\ \cline{2-14}
&\multicolumn{1}{c|}{Pixel} & \multicolumn{1}{c|}{SCT} &\multicolumn{1}{c|}{TRT} & \multicolumn{1}{c|}{LAr} & \multicolumn{1}{c|}{Tile} &\multicolumn{1}{c|}{MDT} & \multicolumn{1}{c|}{RPC}& \multicolumn{1}{c|}{CSC} & \multicolumn{1}{c|}{TGC} &\multicolumn{1}{c|}{Solenoid}& \multicolumn{1}{c|}{Toroid}&\multicolumn{1}{c|}{L1}& \multicolumn{1}{c|}{HLT}\\ \hline\hline
$pp$ @ 13~\TeV  & 98.98 & 99.89 & 99.74 & 99.32 & 99.31 & 99.95 & 99.80 & 100 & 99.96 & 99.15 & 97.23 & 98.33 &100\\
$p$-Pb @ 8.16~\TeV  &  99.92 & 100 & 100 & 100 & 99.99 & 100 &  99.94 & 100 & 100 & 100 & 100 & 100 & 100\\
$p$-Pb @ 5.02~\TeV &    100 & 99.96 & 100 & 100 & 100 & 100 &  99.96 & 100 & 99.95 & 100 & 100 & 100 & 90.44\\\cline{0-13}\noalign{\smallskip}
\cline{5-10}
\multicolumn{1}{c}{}& \multicolumn{3}{c|}{} & \multicolumn{3}{c|}{Data Quality Efficiency [\%]} & \multicolumn{3}{c|}{Integrated Luminosity} \\\cline{0-9}
$pp$ @ 13~\TeV\  & \multicolumn{3}{c|}{\multirow{3}{*}{Good for Physics}}  &  \multicolumn{3}{S[table-format=3.2,table-align-text-post = false,table-number-alignment=center]|}{93.07} & \multicolumn{3}{S[table-format=3.2, table-space-text-post={~fb$^{-1}$}] <{~fb$^{-1}$}|}{33.0}  \\
$p$-Pb @ 8.16~\TeV\ & \multicolumn{3}{c|}{ }  &  \multicolumn{3}{S[table-format=3.2,table-align-text-post = false,table-number-alignment=center]|}{98.35} & \multicolumn{3}{S[table-format=3.2, table-space-text-post={~nb$^{-1}$}] <{~nb$^{-1}$}|}{165}  \\
$p$-Pb @ 5.02~\TeV\ & \multicolumn{3}{c|}{ }  &  \multicolumn{3}{S[table-format=3.2,table-align-text-post = false,table-number-alignment=center]|}{82.93} & \multicolumn{3}{S[table-format=3.2, table-space-text-post={~nb$^{-1}$}] <{~nb$^{-1}$}|}{0.36}  \\ \cline{0-9}
\end{tabular}}
\label{tab:eff-2016}
\end{center}
\end{table}

\begin{table}[h]
\scriptsize
\begin{center}
\caption{Luminosity-weighted relative detector uptime and good data quality efficiencies (in \%)
during stable beams in $pp$ collision physics runs at $\sqrt{s}=13$~\TeV\ between June and November 2017, including 147 pb$^{-1}$ of good $pp$ data taken at an average pile-up of $\mu=2$.
Online luminosity errors, which are not displayed separately in this table, contribute 0.48\% to the total $\sqrt{s}=13$~\TeV\ $pp$ DQ inefficiency.
Also shown are 257~pb$^{-1}$ of good $pp$ data taken at $\sqrt{s}=5.02$~\TeV\ in November
and 1.96~nb$^{-1}$ of Xe-Xe collision data taken at $\sqrt{s_{NN}}=5.44$~\TeV\ in October.
}
\resizebox{\textwidth}{!}{
\begin{tabular}{|l|S[table-format=3.2,table-align-text-post = false,table-number-alignment=center]|S[table-format=3.2,table-align-text-post = false,table-number-alignment=center]|S[table-format=3.2,table-align-text-post = false,table-number-alignment=center]|S[table-format=3.2,table-align-text-post = false,table-number-alignment=center]|S[table-format=3.2,table-align-text-post = false,table-number-alignment=center]|S[table-format=3.2,table-align-text-post = false,table-number-alignment=center]|S[table-format=3.2,table-align-text-post = false,table-number-alignment=center]|S[table-format=3.2,table-align-text-post = false,table-number-alignment=center]|S[table-format=3.2,table-align-text-post = false,table-number-alignment=center]|S[table-format=3.2,table-align-text-post = false,table-number-alignment=center]|S[table-format=3.2,table-align-text-post = false,table-number-alignment=center]|S[table-format=3.2,table-align-text-post = false,table-number-alignment=center]|S[table-format=3.2,table-align-text-post = false,table-number-alignment=center]|}
\hline
\multicolumn{14}{|c|}{\textbf{2017 Data Quality Efficiency [\%]}}  \\ \hline
\multirow{2}{*}{Dataset}&\multicolumn{3}{c|}{Inner Tracker}&\multicolumn{2}{c|}{Calorimeters}&\multicolumn{4}{c|}{Muon Spectrometer}&\multicolumn{2}{c|}{Magnets}&\multicolumn{2}{c|}{Trigger}\\ \cline{2-14}
&\multicolumn{1}{c|}{Pixel} & \multicolumn{1}{c|}{SCT} &\multicolumn{1}{c|}{TRT} & \multicolumn{1}{c|}{LAr} & \multicolumn{1}{c|}{Tile} &\multicolumn{1}{c|}{MDT} & \multicolumn{1}{c|}{RPC}& \multicolumn{1}{c|}{CSC} & \multicolumn{1}{c|}{TGC} &\multicolumn{1}{c|}{Solenoid}& \multicolumn{1}{c|}{Toroid}&\multicolumn{1}{c|}{L1}& \multicolumn{1}{c|}{HLT}\\ \hline\hline
$pp$ @ 13~\TeV\   & 99.97 & 99.94 & 99.32 & 99.48 & 99.41 & 99.90 &  99.24 & 99.89 & 99.98 & 100 & 99.17 & 99.95 & 99.96\\
Xe-Xe @ 5.44~\TeV  & 100 & 100 & 100 & 99.98 & 100 & 100 &  100 & 100 & 100 & 100 & 100 & 100 & 100\\
$pp$ @ 5.02~\TeV  & 100 & 99.73 & 100 & 99.96 & 99.99 & 95.42 &  99.98 & 100 & 99.99 & 100 & 100 & 100 & 100\\
$pp$ @ 13~\TeV\ ($\mu=2$)   &   100  &  100  & 100 &  100 &  100  &  99.49  &  100 &  100 &  100 &    100 &  100 &  100  & 100\\\cline{0-13}\noalign{\smallskip}
\cline{5-10}
\multicolumn{1}{c}{}& \multicolumn{3}{c|}{} & \multicolumn{3}{c|}{Data Quality Efficiency [\%]} & \multicolumn{3}{c|}{Integrated Luminosity} \\\cline{0-9}
$pp$ @ 13~\TeV\            & \multicolumn{3}{c|}{\multirow{4}{*}{Good for Physics}}   &   \multicolumn{3}{S[table-format=3.2,table-align-text-post = false,table-number-alignment=center]|}{95.67} & \multicolumn{3}{S[table-format=3.2, table-space-text-post={~fb$^{-1}$}] <{~fb$^{-1}$}|}{44.39}  \\
Xe-Xe @ 5.44~\TeV                     & \multicolumn{3}{c|}{ } & \multicolumn{3}{S[table-format=3.2,table-align-text-post = false,table-number-alignment=center]|}{99.98} & \multicolumn{3}{S[table-format=3.2, table-space-text-post={~nb$^{-1}$}] <{~nb$^{-1}$}|}{1.96}  \\
$pp$ @ 5.02~\TeV\             & \multicolumn{3}{c|}{ } & \multicolumn{3}{S[table-format=3.2,table-align-text-post = false,table-number-alignment=center]|}{95.20} & \multicolumn{3}{S[table-format=3.2, table-space-text-post={~pb$^{-1}$}] <{~pb$^{-1}$}|}{257} \\
$pp$ @ 13~\TeV\ ($\mu=2$)  & \multicolumn{3}{c|}{ } & \multicolumn{3}{S[table-format=3.2,table-align-text-post = false,table-number-alignment=center]|}{99.28} & \multicolumn{3}{S[table-format=3.2, table-space-text-post={~pb$^{-1}$}] <{~pb$^{-1}$}|}{147}  \\
\cline{0-9}
\end{tabular}}
\label{tab:eff-2017}
\end{center}
\end{table}

\begin{table}[h]
\scriptsize
\begin{center}
\caption{Luminosity-weighted relative detector uptime and good data quality efficiencies (in \%)
during stable beams in $pp$ collision physics runs at $\sqrt{s}=13$~\TeV\ between April and October 2018,
including 193 pb$^{-1}$ of good data taken at an average pile-up of $\mu=2$.
Dedicated luminosity calibration activities during LHC fills used 0.64\% of the recorded data (including $\sim$$4$\% of the $\mu=2$ dataset) and are included in the inefficiency.
Also shown are 1.44~nb$^{-1}$ of good Pb-Pb collision data taken at $\sqrt{s_{NN}}=5.02$~\TeV\ between November and December.
}
\resizebox{\textwidth}{!}{
\begin{tabular}{|l|S[table-format=3.2,table-align-text-post = false,table-number-alignment=center]|S[table-format=3.2,table-align-text-post = false,table-number-alignment=center]|S[table-format=3.2,table-align-text-post = false,table-number-alignment=center]|S[table-format=3.2,table-align-text-post = false,table-number-alignment=center]|S[table-format=3.2,table-align-text-post = false,table-number-alignment=center]|S[table-format=3.2,table-align-text-post = false,table-number-alignment=center]|S[table-format=3.2,table-align-text-post = false,table-number-alignment=center]|S[table-format=3.2,table-align-text-post = false,table-number-alignment=center]|S[table-format=3.2,table-align-text-post = false,table-number-alignment=center]|S[table-format=3.2,table-align-text-post = false,table-number-alignment=center]|S[table-format=3.2,table-align-text-post = false,table-number-alignment=center]|S[table-format=3.2,table-align-text-post = false,table-number-alignment=center]|S[table-format=3.2,table-align-text-post = false,table-number-alignment=center]|}
\hline
\multicolumn{14}{|c|}{\textbf{2018 Data Quality Efficiency [\%]}}  \\ \hline
\multirow{2}{*}{Dataset}&\multicolumn{3}{c|}{Inner Tracker}&\multicolumn{2}{c|}{Calorimeters}&\multicolumn{4}{c|}{Muon Spectrometer}&\multicolumn{2}{c|}{Magnets}&\multicolumn{2}{c|}{Trigger}\\ \cline{2-14}
&\multicolumn{1}{c|}{Pixel} & \multicolumn{1}{c|}{SCT} &\multicolumn{1}{c|}{TRT} & \multicolumn{1}{c|}{LAr} & \multicolumn{1}{c|}{Tile} &\multicolumn{1}{c|}{MDT} & \multicolumn{1}{c|}{RPC}& \multicolumn{1}{c|}{CSC} & \multicolumn{1}{c|}{TGC} &\multicolumn{1}{c|}{Solenoid}& \multicolumn{1}{c|}{Toroid}&\multicolumn{1}{c|}{L1}& \multicolumn{1}{c|}{HLT}\\ \hline\hline
$pp$ @ 13~\TeV\  & 99.78 & 99.77 & 100 & 99.67 &  100 & 99.80 & 99.72 & 99.98 & 99.98 & 100 & 99.58 & 99.99 & 99.99\\
$pp$ @ 13~\TeV\ ($\mu=2$)   & 100  &  100  & 100 &  100  &  100  &  99.07  &  99.94  &  100 &  99.98 &    100 &  100 &  98.03  & 100\\
Pb-Pb @ 5.02~\TeV\  &  100  &  100  & 100 &  99.99  &  100  &  100  &  100  &  100 &  99.98   &    100 &  83.25 &  99.97  & 100\\\cline{0-13}\noalign{\smallskip}
\cline{5-10}
\multicolumn{1}{c}{}& \multicolumn{3}{c|}{} & \multicolumn{3}{c|}{Data Quality Efficiency [\%]} & \multicolumn{3}{c|}{Integrated Luminosity} \\\cline{0-9}
$pp$ @ 13~\TeV\           &   \multicolumn{3}{c|}{\multirow{3}{*}{Good for Physics}}  &  \multicolumn{3}{S[table-format=3.2,table-align-text-post = false,table-number-alignment=center]|}{97.46} & \multicolumn{3}{S[table-format=3.2, table-space-text-post={~fb$^{-1}$}] <{~fb$^{-1}$}|}{58.5}  \\
$pp$ @ 13~\TeV\ ($\mu=2$) &   \multicolumn{3}{c|}{ }                                 &  \multicolumn{3}{S[table-format=3.2,table-align-text-post = false,table-number-alignment=center]|}{92.86} & \multicolumn{3}{S[table-format=3.2, table-space-text-post={~pb$^{-1}$}] <{~pb$^{-1}$}|}{193}  \\
Pb-Pb @ 5.02~\TeV\                    &   \multicolumn{3}{c|}{ }                                 &  \multicolumn{3}{S[table-format=3.2,table-align-text-post = false,table-number-alignment=center]|}{82.54} & \multicolumn{3}{S[table-format=3.2, table-space-text-post={~nb$^{-1}$}] <{~nb$^{-1}$}|}{1.44}  \\
\cline{0-9}
\end{tabular}}
\label{tab:eff-2018}
\end{center}
\end{table}

\begin{table}[h]
\scriptsize
\begin{center}
\caption{Luminosity-weighted relative detector uptime and good data quality efficiencies (in \%)
during stable beams in standard 25~ns $pp$ collision physics runs at $\sqrt{s}=13$~\TeV\ between July 2015 and October 2018.
Dedicated luminosity calibration activities during LHC fills used 0.64\% of the data recorded during 2018, and are included in the inefficiency.
Online luminosity errors, which are not displayed separately in this table, are also included and contribute 0.15\% to the total Run~2 DQ inefficiency.
}
\resizebox{\textwidth}{!}{
\begin{tabular}{|l|S[table-format=3.2,table-align-text-post = false,table-number-alignment=center]|S[table-format=3.2,table-align-text-post = false,table-number-alignment=center]|S[table-format=3.2,table-align-text-post = false,table-number-alignment=center]|S[table-format=3.2,table-align-text-post = false,table-number-alignment=center]|S[table-format=3.2,table-align-text-post = false,table-number-alignment=center]|S[table-format=3.2,table-align-text-post = false,table-number-alignment=center]|S[table-format=3.2,table-align-text-post = false,table-number-alignment=center]|S[table-format=3.2,table-align-text-post = false,table-number-alignment=center]|S[table-format=3.2,table-align-text-post = false,table-number-alignment=center]|S[table-format=3.2,table-align-text-post = false,table-number-alignment=center]|S[table-format=3.2,table-align-text-post = false,table-number-alignment=center]|S[table-format=3.2,table-align-text-post = false,table-number-alignment=center]|S[table-format=3.2,table-align-text-post = false,table-number-alignment=center]|}
\hline
\multicolumn{14}{|c|}{\textbf{Run 2 Data Quality Efficiency [\%]}}  \\ \hline
\multirow{2}{*}{Dataset}&\multicolumn{3}{c|}{Inner Tracker}&\multicolumn{2}{c|}{Calorimeters}&\multicolumn{4}{c|}{Muon Spectrometer}&\multicolumn{2}{c|}{Magnets}&\multicolumn{2}{c|}{Trigger}\\ \cline{2-14}
&\multicolumn{1}{c|}{Pixel} & \multicolumn{1}{c|}{SCT} &\multicolumn{1}{c|}{TRT} & \multicolumn{1}{c|}{LAr} & \multicolumn{1}{c|}{Tile} &\multicolumn{1}{c|}{MDT} & \multicolumn{1}{c|}{RPC}& \multicolumn{1}{c|}{CSC} & \multicolumn{1}{c|}{TGC} &\multicolumn{1}{c|}{Solenoid}& \multicolumn{1}{c|}{Toroid}&\multicolumn{1}{c|}{L1}& \multicolumn{1}{c|}{HLT}\\ \hline\hline
Standard $pp$ @ 13~\TeV\ & 99.50 & 99.85 &99.68&99.52& 99.65 & 99.83 & 99.60 & 99.96& 99.98  &  99.79& 98.84& 99.57 &99.94\\\cline{0-13}\noalign{\smallskip}
\cline{5-10}
\multicolumn{1}{c}{}& \multicolumn{3}{c|}{} & \multicolumn{3}{c|}{Data Quality Efficiency [\%]} & \multicolumn{3}{c|}{Integrated Luminosity} \\\cline{0-9}
Standard $pp$ @ 13~\TeV   &  \multicolumn{3}{c|}{Good for Physics}  &  \multicolumn{3}{S[table-format=3.2,table-align-text-post = false,table-number-alignment=center]|}{\textbf{95.60}} & \multicolumn{3}{S[table-format=3.2,table-align-text-post = false,table-number-alignment=center]|}{\textbf{139.04~fb$^{-1}$}}  \\
\cline{0-9}
\end{tabular}}
\label{tab:eff-run2}
\end{center}
\end{table}

\begin{figure}[htb]
\begin{center}
\includegraphics[width=0.7\linewidth]{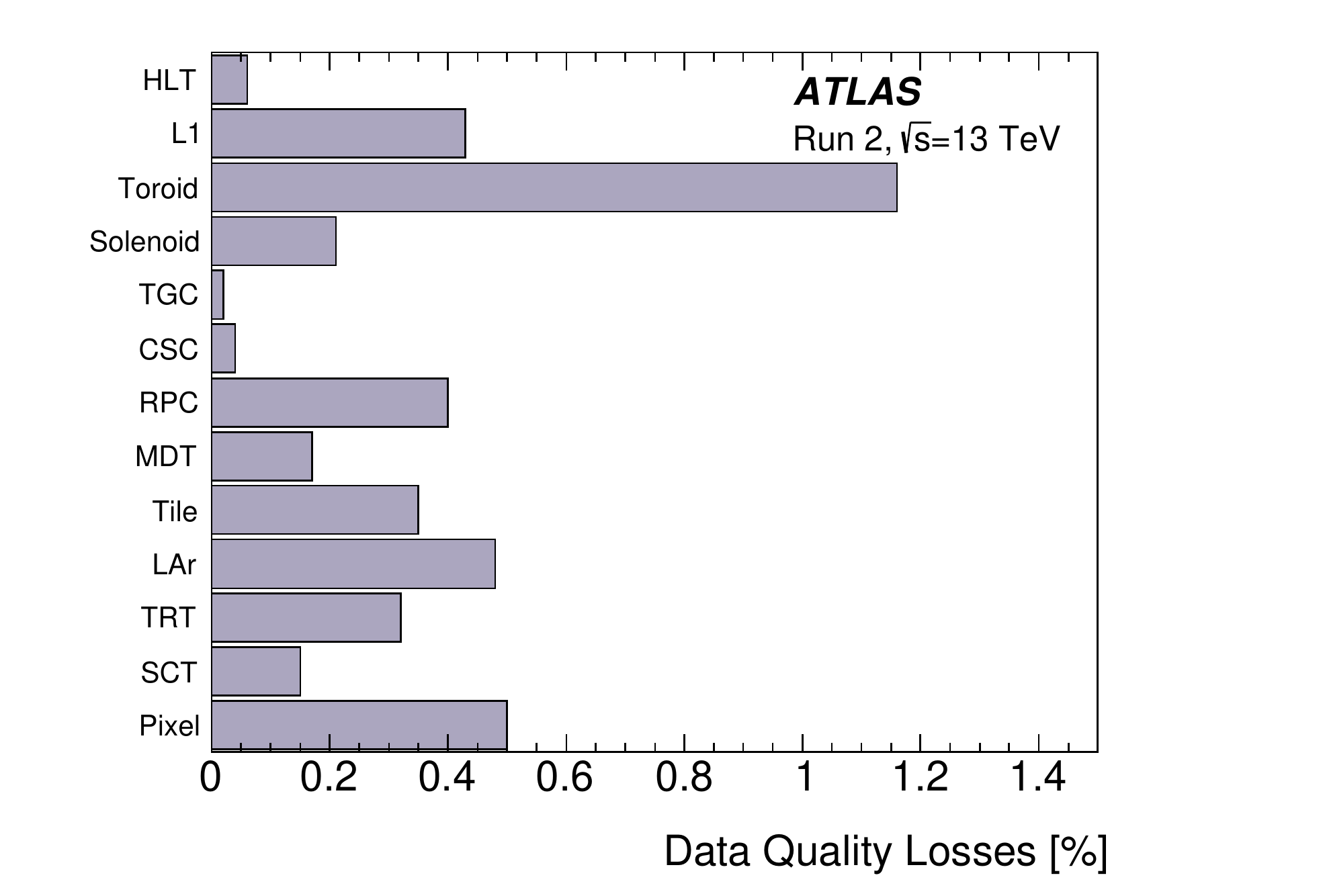}
\caption{Luminosity-weighted data quality inefficiencies (in \%)
during stable beams in standard $pp$ collision physics runs at $\sqrt{s}=13$~\TeV\ between 2015 and 2018.
}
\label{fig:run2_eff}
\end{center}
\end{figure}
 
\begin{figure}[htb]
\begin{center}
\includegraphics[width=0.72\linewidth]{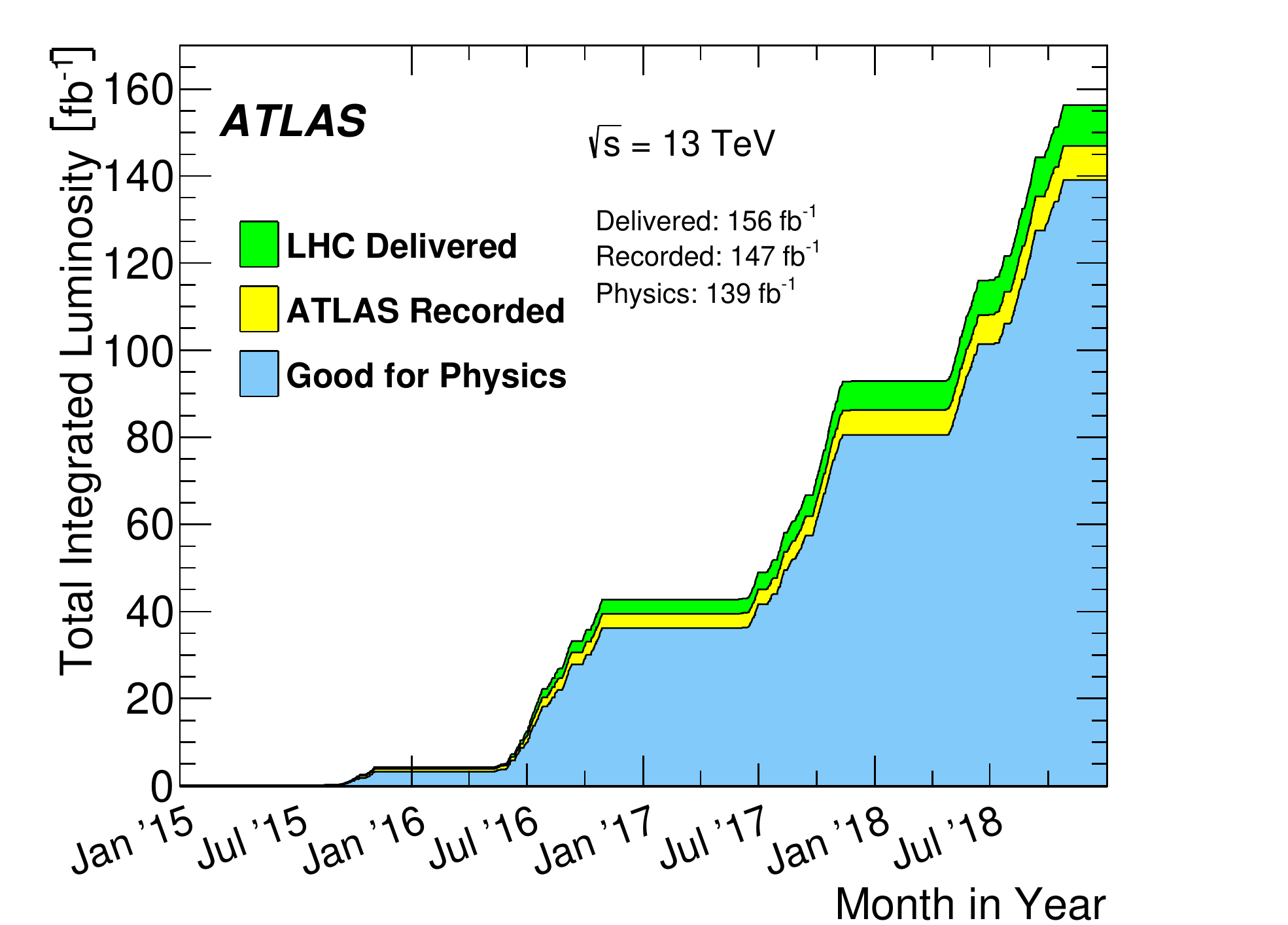}
\includegraphics[width=0.72\linewidth]{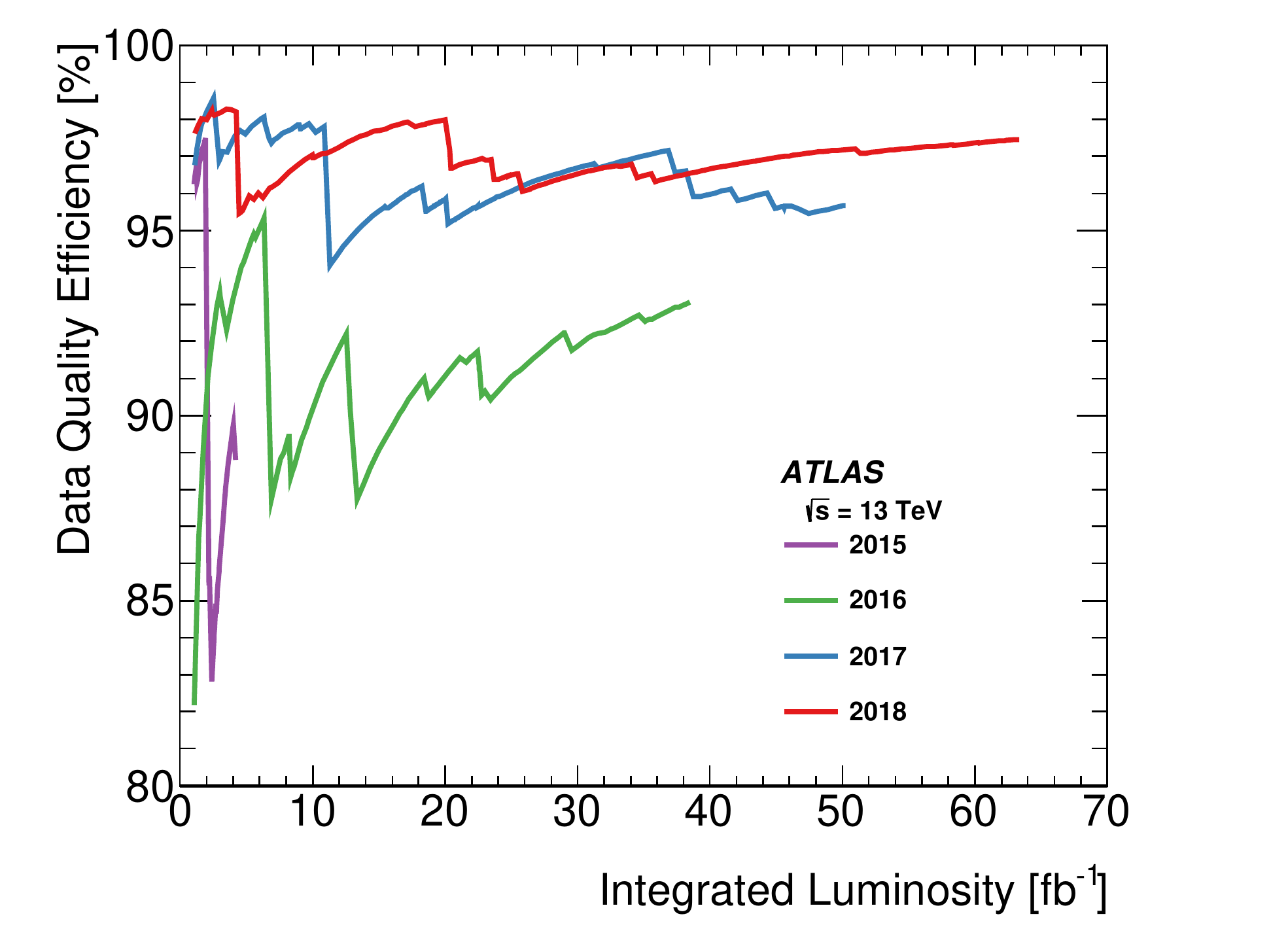}
\caption{Top: Cumulative integrated luminosity delivered to and
recorded by ATLAS between 2015 and 2018 during stable beam $pp$ collision data-taking at $\sqrt{s}=13$~\TeV.
This includes machine commissioning periods, special runs for detector calibration, and
LHC fills with a low number of circulating bunches or bunch spacing greater than 25~ns.
Also shown is the cumulative integrated luminosity certified for physics analysis usage for the ATLAS experiment between 2015 and 2018 during standard $pp$ collision data-taking at $\sqrt{s}=13$~\TeV.
The total integrated luminosity recorded for the standard $\sqrt{s}=13$~\TeV\ $pp$ collision dataset corresponds to 145~fb$^{-1}$.
It is this number that is used in the denominator when calculating the data quality efficiency of the standard $\sqrt{s}=13$~\TeV\ $pp$ collision dataset.
Bottom: Cumulative data quality efficiency versus total integrated luminosity delivered to the ATLAS experiment between 2015 and 2018. }
\label{fig:intlumivstimeRun2DQall}
\end{center}
\end{figure}

The detector subsystems, the L1 trigger, and the HLT achieved excellent DQ efficiencies during Run~2.
While there typically exist known sources of low-level DQ-related data loss at the subsystem or trigger level throughout Run~2 data-taking,
often the largest contributions to the DQ inefficiency come from one-off incidents.
In the case of low-level recurring issues, improved understanding and continuous infrastructure development can mitigate the impact of these sources of DQ losses over time.
For single incidents that result in more-significant DQ losses, lessons must be learned and countermeasures put in place to minimise the risk of similar incidents occurring again.
The main sources of DQ-related losses per subsystem are discussed in the following.
 
The pixel detector began operation in 2015 with the newly installed IBL.
During 2015 data-taking, instabilities were observed from the $\mathrm{CO}_{\mathrm{2}}$ cooling required for IBL operation.
In order to correct the issue it was necessary to take data with the IBL powered off so that the cooling parameters could be optimised.
The 0.2~fb$^{-1}$ of data rejected during this time as a consequence of operating without the IBL constitutes a DQ inefficiency of 6\%,
making up the largest single contribution to the DQ inefficiency for 2015 (see Table~\ref{tab:eff-2015}).
Desynchronisation of data read-out from the pixel detector was a constant source of DQ-related losses in 2015--2017,
but the situation improved over the course of data-taking as a result of continuous firmware and software development, and the replacement of the read-out hardware.
 
Problems related to read-out in the SCT resulted in DQ inefficiencies at the sub-percent level during 2015.
Improvements in the read-out firmware at the end of 2015 resolved this issue, with SCT read-out problems having negligible impact on DQ in the subsequent years.
The largest DQ-related loss for the SCT occurred as a result of a one-time online-software problem during 2018,
with $\sim$$130$~pb$^{-1}$ of $\sqrt{s}=13$~\TeV\ $pp$ data affected.
The software problem caused an immediate halt to SCT clustering, preventing the reconstruction of ID tracks.
Online shifters in the control room identified the absence of tracks using the DQ monitoring displays,
but it took two hours to locate the issue and apply a correction to the relevant software.
Following the software fix,
new monitoring histograms were added to the online monitoring displays to make it easier to quickly identify SCT issues specific to clustering and track reconstruction.
 
The lowest DQ efficiency quoted for the TRT subsystem during Run~2 is shown in Table~\ref{tab:eff-2015} for the 2015 standard $\sqrt{s}=13$~\TeV\ $pp$ dataset, at 98.3\%.
At 1.4\%, most of this inefficiency is due to a single event during November 2015,
where a problematic database update resulted in incorrect latency settings for the TRT being loaded into the online database.
Online monitoring histograms showing the number of TRT hits per track in the control room clearly conveyed that there was a TRT issue,
with the shapes of the observed distributions differing significantly from the reference histograms.
Following expert intervention the issue was resolved, with 50~pb$^{-1}$ of data having been lost.
Database update protocols for the TRT subsystem were improved as a consequence of this incident.
 
High-voltage trips were the major source of DQ losses for the LAr calorimeter system during 2015,
contributing  0.2\% to the total 0.5\% DQ inefficiency quoted for the LAr system during standard $\sqrt{s}=13$~\TeV\ $pp$ data-taking in 2015.
The subsequent replacement of problematic modules with more-robust current control modules at the end of 2015 reduced the impact of HV trips from 2016 onwards by over 90\%.
The largest DQ-related loss for both the LAr and Tile calorimeter subsystems during 2016 resulted from the same incident, described in the following.
For a single run in 2016, a misconfiguration in the online database resulted in calorimeter cell noise thresholds being too low.
This resulted in problems for the trigger, with too many clusters being created from the calorimeter, overloading the HLT computing farm.
Following intervention by software experts, the situation was resolved but,
in the meantime, emergency measures had to be put into effect by the trigger, making the data collected effectively unusable for physics analyses.
In total, $\sim$$140$~pb$^{-1}$ of $pp$ data were rejected as a result of this issue,
contributing 0.4\% to both the total 0.7\% LAr inefficiency and the total 0.7\% Tile inefficiency during standard $\sqrt{s}=13$~\TeV\ $pp$ data-taking quoted for 2016 in Table~\ref{tab:eff-2016}.
As a result of this incident,
software was introduced to allow trigger experts to run tests after an update is made to the online database to validate the database configuration prior to the start of a new run.
In addition, calorimeter cluster occupancy monitoring was improved, with dedicated histograms being introduced at the central DQ desk in the control room to be monitored by the DQ shifter.
Hardware issues resulting in an intolerable loss of detector coverage also affected both the LAr and Tile calorimeter subsystems during 2017.
At the end of August, $\sim$$120$~pb$^{-1}$ of data were lost when a water leak resulted in a power supply trip,
disabling a quarter of the LAr barrel calorimeter.
A hardware issue also resulted in half of the Tile barrel being disabled,
with a subsequent data loss of $265$~pb$^{-1}$ of $\sqrt{s}=13$~\TeV\ $pp$ collision data.
Following this incident the affected hardware was inspected and, where necessary, replaced, to avoid future occurrences of similar problems.
 
The muon system operated with excellent DQ performance during Run~2, with a DQ efficiency consistently at or above $99.8$\%.
An exception to this was observed in 2017 for the RPC system, where a hardware problem resulted in an unacceptable coverage loss on both sides of the detector.
The extent of the loss in coverage is shown in Figure~\ref{fig:rpc}, where the absence of hits being recorded in individual RPC sectors indicates the problem.
The issue was identified in real time in the control room due to DCS alarms indicating that the RPC currents were not as expected,
and a hardware intervention was necessary to replace a problematic HV crate.
Given the impact of this loss in coverage for muon reconstruction, as shown in the bottom panel of Figure~\ref{fig:rpc},
this single event caused the rejection of $\sim$$500$~pb$^{-1}$ of $pp$ collision data, resulting in a lower than usual DQ efficiency of $99.2$\% for the RPC subsystem in 2017.
 
\begin{figure}[htb]
\begin{center}
\includegraphics[width=0.7\linewidth]{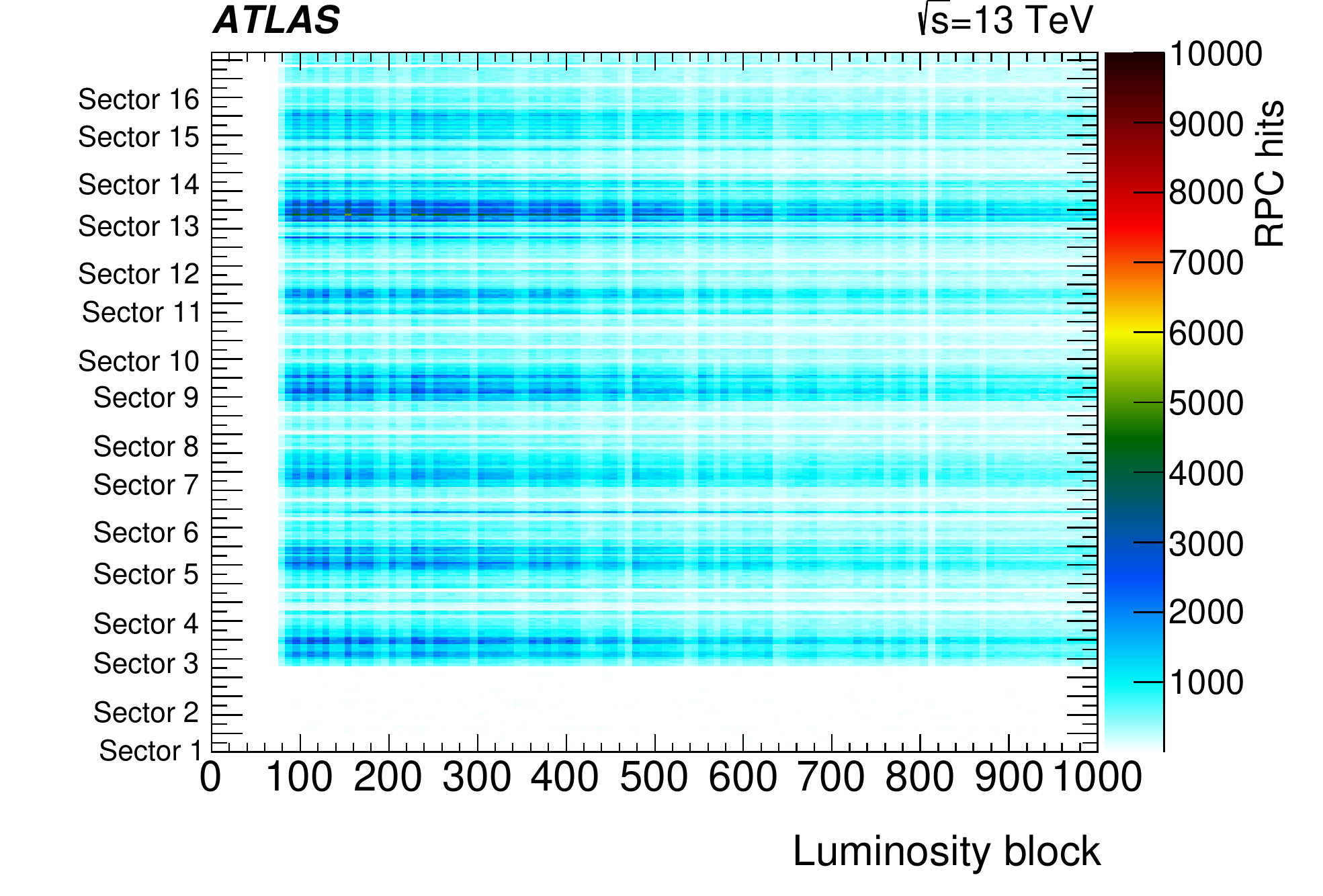}
\includegraphics[width=0.7\linewidth]{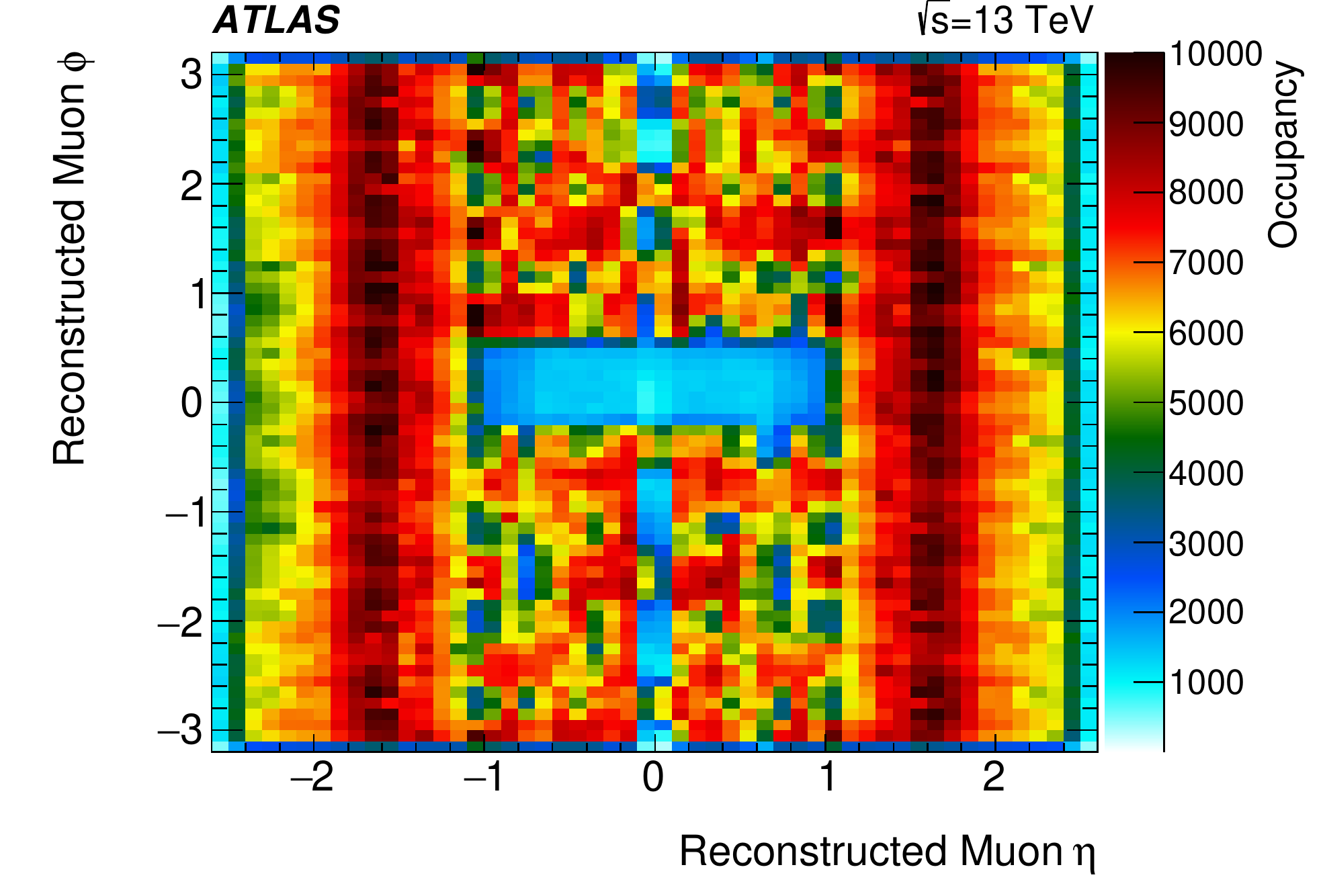}
\caption{Illustration of an incident during 2017, which resulted in a significant loss of coverage for muons.
Top: Number of hits in the various sectors of the RPC subsystem versus luminosity block, where the absence of entries for $\phi$-sectors 1 and 2 indicates a hardware issue.
Bottom: Occupancy of reconstructed muons in the $\eta$--$\phi$ plane, where the deficit in the range $-1<\eta<1$ at $\phi\sim0$ illustrates the impact of this hardware problem on the reconstructed objects.}
\label{fig:rpc}
\end{center}
\end{figure}

The L1 and HLT trigger systems generally rely on information provided by detector subsystems, and therefore most trigger-related DQ issues can be traced to a detector problem further upstream.
The trigger operated throughout Run~2 with very little DQ-related loss, as indicated in Table~\ref{tab:eff-run2}.
During 2016, however, an L1 trigger hardware issue resulted in a mis-timing of muon and jet triggers over the course of four runs, affecting $\sim$$590$~pb$^{-1}$ of $pp$ collision data.
The resulting drop in DQ efficiency can be seen in Figure~\ref{fig:intlumivstimeRun2DQall} as a sharp drop in the 2016 DQ efficiency at $\sim$$12$~\ifb.
This is reflected in the L1 DQ efficiency of 98.3\% for the $\sqrt{s}=13$~\TeV\ $pp$ collision dataset in 2016, as shown in Table~\ref{tab:eff-2016}.
 
Data taken during emittance scans are used for luminosity calibration purposes,
but the non-standard vdM-like configuration is unsuitable for standard physics analyses.
These data must therefore be rejected from physics analysis GRLs.
Dedicated emittance scans for luminosity calibration taken during standard data-taking runs accounted for 0.64\% of the total DQ inefficiency during 2018.
 
The beginning of Run~2 data-taking during 2015 involved commissioning new detector hardware such as the IBL,
and required the careful monitoring of all ATLAS subsystems to understand any constraints or potential issues connected with operating under the more challenging machine conditions of Run~2.
A high DQ efficiency was achieved as a consequence of continuous detector maintenance, software development and monitoring of DQ performance over the course of Run~2.
As the underlying sources of DQ losses were identified and addressed, actions were taken to reduce the probability of subsequent occurrences.
By learning from DQ-related issues in this way the overall ATLAS DQ efficiency during standard $\sqrt{s}=13$~\TeV\ $pp$ data-taking steadily increased from 88.8\% to 97.5\% between 2015 and 2018.
This increase can be seen in the bottom panel of Figure~\ref{fig:intlumivstimeRun2DQall}.
\FloatBarrier
\section{Summary}
\label{sec:conclusion}
The ATLAS data quality operating scheme and assessment procedures as performed during LHC Run~2, and the data quality performance achieved as a result, are presented.
The data quality performance for the standard $pp$ collision dataset is presented,
alongside the corresponding efficiency for special running periods, including heavy-ion data-taking and dedicated low-$\mu$ datasets.
Data quality and ATLAS operating procedures have been subject to continuous review and development over the course of the Run~2 data-taking campaign,
resulting in an increasing efficiency for good quality physics data.
A data quality efficiency for the standard $pp$ collision dataset collected at $\sqrt{s}=13$~\TeV\ of 97.5\% was achieved during 2018,
completing a standard Run~2 dataset taken between 2015 and 2018 with a combined data quality efficiency of 95.6\%.
 
\section*{Acknowledgements}
 

We thank CERN for the very successful operation of the LHC, as well as the
support staff from our institutions without whom ATLAS could not be
operated efficiently.
 
We acknowledge the support of ANPCyT, Argentina; YerPhI, Armenia; ARC, Australia; BMWFW and FWF, Austria; ANAS, Azerbaijan; SSTC, Belarus; CNPq and FAPESP, Brazil; NSERC, NRC and CFI, Canada; CERN; CONICYT, Chile; CAS, MOST and NSFC, China; COLCIENCIAS, Colombia; MSMT CR, MPO CR and VSC CR, Czech Republic; DNRF and DNSRC, Denmark; IN2P3-CNRS and CEA-DRF/IRFU, France; SRNSFG, Georgia; BMBF, HGF and MPG, Germany; GSRT, Greece; RGC and Hong Kong SAR, China; ISF and Benoziyo Center, Israel; INFN, Italy; MEXT and JSPS, Japan; CNRST, Morocco; NWO, Netherlands; RCN, Norway; MNiSW and NCN, Poland; FCT, Portugal; MNE/IFA, Romania; MES of Russia and NRC KI, Russia Federation; JINR; MESTD, Serbia; MSSR, Slovakia; ARRS and MIZ\v{S}, Slovenia; DST/NRF, South Africa; MINECO, Spain; SRC and Wallenberg Foundation, Sweden; SERI, SNSF and Cantons of Bern and Geneva, Switzerland; MOST, Taiwan; TAEK, Turkey; STFC, United Kingdom; DOE and NSF, United States of America. In addition, individual groups and members have received support from BCKDF, CANARIE, Compute Canada and CRC, Canada; ERC, ERDF, Horizon 2020, Marie Sk{\l}odowska-Curie Actions and COST, European Union; Investissements d'Avenir Labex, Investissements d'Avenir Idex and ANR, France; DFG and AvH Foundation, Germany; Herakleitos, Thales and Aristeia programmes co-financed by EU-ESF and the Greek NSRF, Greece; BSF-NSF and GIF, Israel; CERCA Programme Generalitat de Catalunya and PROMETEO Programme Generalitat Valenciana, Spain; G\"{o}ran Gustafssons Stiftelse, Sweden; The Royal Society and Leverhulme Trust, United Kingdom.
 
The crucial computing support from all WLCG partners is acknowledged gratefully, in particular from CERN, the ATLAS Tier-1 facilities at TRIUMF (Canada), NDGF (Denmark, Norway, Sweden), CC-IN2P3 (France), KIT/GridKA (Germany), INFN-CNAF (Italy), NL-T1 (Netherlands), PIC (Spain), ASGC (Taiwan), RAL (UK) and BNL (USA), the Tier-2 facilities worldwide and large non-WLCG resource providers. Major contributors of computing resources are listed in Ref.~\cite{ATL-GEN-PUB-2016-002}.
 

\clearpage

\printbibliography

\clearpage 
 
\begin{flushleft}
{\Large The ATLAS Collaboration}

\bigskip

G.~Aad$^\textrm{\scriptsize 102}$,    
B.~Abbott$^\textrm{\scriptsize 129}$,    
D.C.~Abbott$^\textrm{\scriptsize 103}$,    
A.~Abed~Abud$^\textrm{\scriptsize 36}$,    
K.~Abeling$^\textrm{\scriptsize 53}$,    
D.K.~Abhayasinghe$^\textrm{\scriptsize 94}$,    
S.H.~Abidi$^\textrm{\scriptsize 167}$,    
O.S.~AbouZeid$^\textrm{\scriptsize 40}$,    
N.L.~Abraham$^\textrm{\scriptsize 156}$,    
H.~Abramowicz$^\textrm{\scriptsize 161}$
% ...
\bigskip
\\

% ...
$^{36}$CERN, Geneva; Switzerland.\\
% ...
$^{40}$Niels Bohr Institute, University of Copenhagen, Copenhagen; Denmark.\\
% ...
$^{53}$II. Physikalisches Institut, Georg-August-Universit\"{a}t G\"ottingen, G\"ottingen; Germany.\\
% ...
$^{94}$Department of Physics, Royal Holloway University of London, Egham; United Kingdom.\\
% ...
$^{102}$CPPM, Aix-Marseille Universit\'e, CNRS/IN2P3, Marseille; France.\\
$^{103}$Department of Physics, University of Massachusetts, Amherst MA; United States of America.\\
% ...
$^{129}$Homer L. Dodge Department of Physics and Astronomy, University of Oklahoma, Norman OK; United States of America.\\
% ...
$^{156}$Department of Physics and Astronomy, University of Sussex, Brighton; United Kingdom.\\
% ...
$^{161}$Raymond and Beverly Sackler School of Physics and Astronomy, Tel Aviv University, Tel Aviv; Israel.\\
% ...
$^{167}$Department of Physics, University of Toronto, Toronto ON; Canada.\\
% ...

\end{flushleft}


\end{document}